\documentclass[sigconf]{acmart}

\AtBeginDocument{%
  \providecommand\BibTeX{{%
    \normalfont B\kern-0.5em{\scshape i\kern-0.25em b}\kern-0.8em\TeX}}}

\setcopyright{iw3c2w3}
\copyrightyear{2021}
\acmYear{2021}
\acmConference[WWW '21]{Proceedings of the Web Conference 2021}{April 19--23, 2021}{Ljubljana, Slovenia}
\acmBooktitle{Proceedings of the Web Conference 2021 (WWW '21), April 19--23, 2021, Ljubljana, Slovenia}
\acmPrice{}
\acmDOI{10.1145/3442381.3450122}
\acmISBN{978-1-4503-8312-7/21/04}

\usepackage[table]{xcolor}

\usepackage{latexsym}
\usepackage{url}
\usepackage{verbatim}
\usepackage{multirow}
\usepackage{booktabs}
\usepackage{graphicx}
\usepackage{subcaption}
\usepackage{xspace}
\usepackage{makecell}
\usepackage[table-number-alignment=center, group-separator={}]{siunitx}
\def\sym#1{\ifmmode^{#1}\else\(^{#1}\)\fi}
\newcommand{\tlc}{\textsc{Tlc}\xspace}

\newcommand{\Sref}[1]{\S\ref{#1}}
\newcommand{\sref}[1]{\S\ref{#1}}
\newcommand{\fref}[1]{Figure~\ref{#1}}

\newcommand{\mypara}[1]{\paragraph{#1}}

\title{Conversations Gone Alright: Quantifying and Predicting Prosocial Outcomes in Online Conversations}

\newcommand{\newcite}[1]{\citeauthor{#1}}
\usepackage{fontawesome}
\usepackage[nointegrals]{wasysym}
\usepackage{makecell}

\begin{document}

\author{Jiajun Bao}
\authornote{Authors contributed equally to this research.}
\email{jiajunb@andrew.cmu.edu}
\affiliation{%
  \institution{Carnegie Mellon University, Language Technologies Institute}
}

\author{Junjie Wu}
\authornotemark[1]
\email{junjie.wu@connect.ust.hk}
\affiliation{%
  \institution{The Hong Kong University of Science and Technology}
}

\author{Yiming Zhang}
\authornotemark[1]
\email{yimingz@umich.edu}
\affiliation{%
  \institution{University of Michigan}
}

\author{Eshwar Chandrasekharan}
\email{eshwar@illinois.edu}
\affiliation{%
  \institution{University of Illinois, Urbana-Champaign}}

\author{David Jurgens}
\email{jurgens@umich.edu}
\affiliation{%
  \institution{University of Michigan, School of Information}
}

\renewcommand{\shortauthors}{Bao,* Wu,* Zhang,* Chandrasekharan, and Jurgens}

\begin{abstract}

Online conversations can go in many directions: some turn out poorly due to antisocial behavior, while others turn out positively to the benefit of all.  Research on improving online spaces has focused primarily on detecting and reducing antisocial behavior.  Yet we know little about positive outcomes in online conversations and how to increase them---is a prosocial outcome simply the lack of antisocial behavior or something more?  Here, we examine how conversational features lead to prosocial outcomes within online discussions. We introduce a series of new theory-inspired metrics to define prosocial outcomes such as mentoring and esteem enhancement. Using a corpus of 26M Reddit conversations, we show that these outcomes can be forecasted from the initial comment of an online conversation, with the best model providing a relative 24\% improvement over human forecasting performance at ranking conversations for predicted outcome.
Our results indicate that platforms can use these early cues in their algorithmic ranking of early conversations to prioritize better outcomes.
\end{abstract}


\maketitle

\section{Introduction}

Interacting with others online has become a common facet of daily life.
Yet, these interactions often can turn out poorly, in part due to toxic behavior on the part of others \cite{kumar2017antisocial,duggan2017online}. 
Given the significant impact of experiencing these negative activities on well-being \citep{roberts2014behind,gillespie2018custodians,saha2019prevalence}, substantial research effort has been put into detecting such toxic behavior and facilitating platform tools to remove it.
However, despite sophisticated techniques for measuring antisocial behavior, the key metrics for \textit{prosocial} behavior are relatively unknown, to the point that major platforms such as Twitter and Instagram have both called for researchers to develop such metrics \cite{twitter2018health,instagram2019well}.
Here, we operationalize theories from social psychology to quantify and measure prosocial behavior in social media, showing a rich diversity in the types of behaviors, and then show that these positive outcomes can be forecasted from the early stages of a conversation.

\begin{figure}[t]
    \centering
    \begin{tabular}{|p{7.46cm}|}
  
    \hline
     \small
    \textbf{Post}: ``Studied like crazy all last week and finally took a Technician course and tested on Saturday. Boom, I'm legal! 
    Thanks for all of the support I've got on /r/amateurradio!!" \\
  \small $\llcorner$ \textbf{Reply 1}: 
  That test is a killer! \\ 
 \small $\llcorner$ \textbf{Reply 2}: 
 Great job! Hope to hear you on the air some day \\  

    \hline
    \end{tabular}
    
    \caption{Which reply is likely to lead to a positive, productive conversation? Here, we introduce new metrics for measuring the prosocial qualities of social media discussions and develop new models to predict which of these replies will lead to a conversation with higher prosocial behavior. }
    \label{fig:intro-example}
\end{figure}

The impact of online discussions on daily life and mental health has prompted multiple studies on online conversational dynamics.
A significant effort has focused on detecting antisocial behaviors such as hate speech~\cite{waseem2017understanding} , trolling~\cite{cheng2017anyone} , or bullying~\cite{liu2018forecasting}, in attempts to mitigate their effect by offering moderators tools to find and remove them.
Further, recent work has shown that the initial start of a conversation can forecast antisocial outcomes from early linguistic and behavioral information \citep{liu2018forecasting,zhang2018conversations,jiao2018find,chang2019trouble}.
However, only a handful of studies have examined prosocial behaviors, such as making constructive comments \cite{kolhatkar2017constructive,napoles2017automatically,kolhatkar2020classifying} or offering supportive messages  \cite{wang2015pragmatic,navindgi2016steps,wang2018its}.  
Our work brings together these lines of research through a systematic examination of prosocial behaviors and building models to forecast these conversational outcomes.

We introduce eight types of prosocial metrics and develop new methods to forecast the prosocial trajectory of a conversation from its early interactions.  
Focusing this task on one of prediction is strongly motivated by the implications for real-world impact.  
Online platforms regularly engage in content reranking where comments and threads are reordered according to internal objectives \cite{bucher2012want,lazer2015rise}.
Given the ability to predict the prosocial trajectory of a conversation, platforms can potentially rerank the initial comments to a post (or other comments) to emphasize those that will lead to better community experiences.

This paper offers the following three contributions.
Using theoretical insights from prior literature on prosocial and antisocial behavior in online and offline contexts (\sref{sec:prosocial}), we introduce a panel of prosocial metrics and construct a large-scale corpus of social media conversations labeled by these outcomes (\sref{sec:data}). Using this corpus, we demonstrate that these metrics are significantly associated with human judgments of prosociality and show that prosociality is not just the absence of antisocial behavior. 
Second, we introduce four new models for forecasting the prosocial quality of a conversation (\sref{sec:forecasting}), showing that such outcomes can be accurately forecasted from  cues early in the conversation.  
Third, given the first comment of two conversations, we demonstrate that both our models and people can forecast which conversation is likely to turn out better  (\Sref{sec:ranking}), with our model offering a 24\% improvement on human accuracy with respect to chance.
Our work has implications for platforms' abilities to  surface online interactions in order to create positive outcomes for individuals participating in them.

\section{Prosocial Behavior}
\label{sec:prosocial}

Prosocial behavior began as an antonym used by social scientists for describing the opposite of antisocial behavior \cite{knickerbocker2003prosocial,bar1976prosocial}, with \newcite{mussen1977roots} defining the behavior  as  ``voluntary actions that are intended to help or benefit another individual or group of individuals.''  Since this time, prosocial behavior has broadened to include a range of activities: helping, sharing, comforting, rescuing, and cooperating \cite{batson2003altruism}.  
Our work examines prosocial behavior in online discussions by deriving a large cohort of candidate metrics for measuring conversations from theory and then testing which are associated with judgments of prosocial behavior online.

The concept of prosociality is complex and the nuances of which aspects of behavior contribute to its perception online are not yet well understood. 
A few recent approaches examining specific factors related to positive conversational outcomes like constructive comments \cite{napoles2017automatically,napoles2017finding}, politeness \cite{danescu2013computational}, supportiveness \cite{wang2018its}, or empathy  \cite{zhou2020condolences,buechel2018modeling,sharma2020computational};
or, showing that, in general, online prosocial behaviors mirror offline trends \cite{wright2011associations}.
In the majority of cases, only individual dimensions have been analyzed; however, we note that recent work has proposed studying these dimensions jointly in relationships and social interactions \cite{choi2020ten} using the ten social dimensions outlined in \citet{deri2018coloring}. Similar to the present work, \citet{choi2020ten} examines general factors from sociological and psychological literature for relationships to study interactions; however, the factors used here are specifically grounded in prosocial literature and include behavioral factors in addition to linguistic factors.
A few studies have tried to measure prosocial behavior as a single variable \cite{frimer2014moral,frimer2015decline}; however, these approaches in practice have used lexicons that recognize only a subset of the possible prosocial behaviors focused on collective interest and interpersonal harmony. 
 

Prosocial behaviors can take many forms depending on the parties involved and their needs. Here, we identify eight broad categories of behavior ground in prior work from Social Psychology that can be easily operationalized using NLP techniques and are markers of direct prosocial behavior or are behaviors that serve as a precursor to prosocial behavior. As many of these behaviors have been identified and studied in offline settings, our aim is to study how these behaviors are interpreted in online settings in order to curate a set of prosocial metrics that match peoples' readings of online interactions. Following, we outline each category, its connection to prior theory, and summarize how its behavior is recognized. Additional details for metrics and classifiers are provided in Appendices \ref{app:info-share-sec}--\ref{app:support}.

\mypara{Information Sharing} 
Individuals seek out information online where others may provide suggestions.  In some settings these efforts are codified around collaboratively creating information goods like Wikipedia or open source projects \cite{sproull2011prosocial}. In social media like Reddit, responses to questions create persistent knowledge that can be learned from by others. This knowledge transfer may take the form of explicit information or references to websites such as Wikipedia or StackOverflow. Here, we operationalize these information sharing behaviors in two ways. First, using information-providing based subreddits (e.g., \texttt{r/AskScience}), we train a classifier to recognize informative replies to questions; and then as a prosocial metric, use the classifier to identify and count these replies in a conversation---i.e., does a discussion lead to information sharing? Second, recognizing that URLs often serve as important sources of third-party information, we include counts for (i) information-based domains, e.g., wikipedia.org, and (ii) for all other websites, recognizing that many domains may serve informative purposes (e.g., linking to a product review). More details about the training of the classifier can be found in Appendix \ref{app:info-share-sec}.

\mypara{Gratitude}
Gratitude serves an important function in fostering social relationships and promoting future reciprocity \cite{emmons2004psychology,bartlett2006gratitude}.  Gratitude not only reinforces existing prosocial behavior, but also motivates more prosocial behavior and itself serves as an indicator that prosocial behavior has occurred \cite{mccullough2001gratitude,mccullough2008adaptation,algoe2012find}. Here, we identify gratitude through a fixed lexicon of phrases that signal gratitude (Appendix \ref{appendix-gratitude}), e.g., ``thank you,'' and count how many times such phrases are used in a discussion.

\mypara{Esteem Enhancement}
Prosocial behavior is known to be motivated by a person's self-esteem \cite{batson2003altruism}. Individuals may seek out opportunities to behave prosocially as a way to repair or improve their self-esteem \cite{brown1991self}; improved esteem can increase the perception of reciprocity for help, further motivating prosocial actions. Thus, esteem-enhancing actions can serve as a useful behavior to monitor as precursors for prosocial behavior.
Here, we measure esteem enhancement using three metrics. First, recognizing that politeness is often used to signal social status  \cite{brown2015politeness}, we use a politeness regressor based in part on \citet{danescu2013computational} to measure the average politeness of comment interactions (Appendix \ref{appendix-politeness}); the underlying hypothesis is that more polite messages increase the status (esteem) perception of the recipient. 
Second, we identify all statements with second-person pronominal references (e.g., you) and count how many have strongly positive  sentiment, which approximates identifying compliments (Appendix \ref{appendix-esteem}). 
Third, we measure the total score given to responses in a discussion as a measure of esteem given by the community to the conversation. The score of a comment is closely correlated with the number of upvotes it receives (as derived through a proprietary measure)  and receiving upvotes is known to be an esteem enhancing action \cite{burrow2017many}.

\begin{table*}[t!]
    \centering
    \small
    \resizebox{\textwidth}{!}{
    \begin{tabular}{l p{200pt} S[table-align-text-post=false,table-format=1.4] c}
        \textbf{Prosocial Metric}& \textbf{Description}& {\textbf{MCC}} &
        \textbf{Percentage}\\ 
        \hline

        



        %
        %
        
        

        
        
        %
        %

        \cellcolor{purple!10} Information sharing & number of replies classified as informative & 0.3452*** & 0.1218\\ 

        \cellcolor{purple!10} Link replies & number of links (urls) in all replies & 0.5129*** & 0.1095\\ 
        
        \cellcolor{purple!10} Educational Link replies & number of educational links (urls) in all replies & 0.6237*** & 0.0070\\ 

        %
        %
  
        \cellcolor{green!10} Gratitude & number of gratitude in all replies & 0.3154*** & 0.1096\\ 

        %
        %
        
        \cellcolor{orange!10} Politeness & average politeness value of all replies & 0.0346 & 1.0000\\ 
        
        \cellcolor{orange!10} Linguistic accommodation  & Replies to the \tlc mirror function word usage  & 0.0997$\dagger$ & 0.5208\\ 

        \cellcolor{orange!10} Community score  & numeric sum of every reply's score & 0.5531*** & 1.0000\\ 

        %
        %
        
        \cellcolor{black!05}  Supportiveness  & average supportiveness value of all replies & 0.0511 & 1.0000\\ 
        

        %
        
        %

        \cellcolor{green!20} Subsequent comments & a top comment's total number of replies & 0.5988*** & 1.0000\\ 
        
        \cellcolor{green!20} Direct replies & number of replies responding directly to the \tlc & 0.1529*** & 1.0000\\ 
        
        \cellcolor{green!20} Conversation depth & length of the longest replies' thread & 0.5400*** & 1.0000\\ 

        \cellcolor{green!20} Sustained conversation partners & number of distinct user pairs appear in all replies & 0.5508*** & 0.4351 \\ 
        
        \cellcolor{green!20} Sustained conversations & number of turns in the longest two-person conversation & 0.6212*** & 0.4351\\ 
        

        \cellcolor{green!20} Compliments & number of compliments & 0.5374*** & 0.0045\\      
        
        \cellcolor{green!20} Laughter & number of expressions of laughter & 0.2678*** & 0.0716\\

        \cellcolor{green!20} Personal disclosure & number of statements an authors makes about themself & 0.4091*** & 0.3302\\ 

        %
        %
        
        \cellcolor{purple!20} Donations & number of links to charities and donation sites & 0.2987 & 0.0002\\ 
        
        \cline{1-1}        
        %
        %
        
        \cellcolor{yellow!10} Mentorship  & number replies classified as mentoring & 0.3957*** & 0.0950\\ 

        

        \cellcolor{blue!15} $\%$ of non-toxicity (untuned) & percentage of non toxic replies in all replies & 0.3363*** & 0.1392\\ 
        
        \cellcolor{blue!15} $\%$ of non-toxicity (tuned) & percentage of non extremely toxic replies in all replies & 0.1222 & 0.0239\\ 
        
        \cellcolor{blue!15} Untuned toxic language  & number of toxic replies & -0.2852*** & 0.1392\\ 
        
        \cellcolor{blue!15} Tuned toxic language & number of extremely toxic replies & -0.1014 & 0.0239\\ 

        \hline
    \end{tabular}
    }
    \caption{Theory-based metrics used to measure prosocial outcomes in conversations (colored by category) and their correlation with human judgments of prosociality, using Matthew's Correlation Coefficient (MCC; see \Sref{sec:prosocial-judgments} for details) and their rates of occurrence in conversations. 
    Throughout the paper, we use *** to denote p$<$0.001, ** p$<$0.01, and * p$<$0.05, here shown after Bonferroni correction. $\dagger$Acommodation varied in significance across all types, with 10 having a significant MCC (shown here with the mean).} 
    \label{tab:metrics-table}
\end{table*}

\mypara{Social Support}
In times of distress, individuals turn to their social network for support \cite{wills1991social,batson2003altruism}. Online communities and platforms have provided a parallel support mechanism around many types of needs, such as physical and mental health \cite{biyani2014identifying,wang2015pragmatic,milne2016clpsych} and weight-loss. Moreover, beyond specific needs, individuals  offer supportive messages in general on these platforms, e.g., encouragement \cite{wang2018its}. 
Due to the diversity of support expected on Reddit, we develop a computational model to recognize supportive messages using the data of \newcite{wang2018its} (Appendix \ref{app:support}) and use the average supportiveness of comments in a conversation. 

\mypara{Social Cohesion}
Social ties create a sense of community, which carries with it the benefits of group membership and altruistic behavior between members \cite{akerlof2000economics,goette2012impact}.  Conversely, exclusions from a group or a weakening of ties decrease prosocial behavior \cite{twenge2007social}.   Indeed,  \newcite{prinstein2003forms} note that helping someone join a conversation is a core prosocial behavior and studies have shown that increased linguistic accommodation \cite{niederhoffer2002linguistic,taylor2008linguistic} is associated with increased prosocial behavior \cite{kulesza2014echo} and trust \cite{scissors2008linguistic}.

To measure the formation of social bonds, we use four categories of metrics. First, building on the insight of \citet{kulesza2014echo}, we measure linguistic accommodation between commenters using the methods of \newcite{danescu2012echoes}. 
Second, recognizing that increased conversation gives rise to social ties, we include metrics for (i) the total number of participants in a conversation, (ii) the longest number of sustained turns between two people, and (iii) the depth of the conversation's comment tree.
Third, laughter, as a function of humor, is known to create positive affect between peers and increase cohesion among group members \cite{bachorowski2001not,owren2003reconsidering,greatbatch2003displaying}. Therefore, we count the number of laughter events in a conversation (see Appendix \ref{appendix-laughter} for details). 
Fourth, self disclosure is known to strengthen social ties \cite{duck2007human,joward1971self,joinson2007self}; to measure disclosures, we follow prior work \cite{joinson2001self,barak2007degree,bak2014self} and count the number of comments including a first person pronoun.

\mypara{Fundraising and Donating}
An offline prosocial behavior that readily transfers to online behavior is fundraising for charitable activities \cite{sproull2011prosocial,wright2012prosocial}. Many online charities have websites set up to receive donations and sites such as gofundme for more individualized fundraising activities. Here, we count the number of URLs to these types of sites using a  list of popular charities, detailed in Appendix \ref{appendix-donating}.

\mypara{Mentoring}
Individuals can give expertise through mentoring or advice when others are having a problem, which is a known prosocial behavior \cite{prinstein2003forms,wright2012prosocial}.
To recognize advice giving and mentoring, we train a classifier to distinguish the language of advice in the responses of advice-based subreddits (e.g., \texttt{\small /r/FashionAdvice}, \texttt{\small /r/RelationshipAdvice}, \texttt{\small /r/LegalAdvice}) from responses in other subreddits. Then, we use this classifier to recognize and count the number of advice-based responses in a conversation thread. Further details on the classifier can be found at \ref{appendix-mentoring}.

\mypara{Absence of Antisocial Behavior}
Is prosocial behavior implied by the absence of antisocial behavior?  If true, this question suggests that platforms' efforts to find and remove antisocial behavior are sufficient for retaining and promoting prosocial behavior.
To answer this question, we label all replies using the Perspective API \cite{wulczyn2017ex}, which assigns a score for toxicity.  Highly toxic content like explicit hate speech is assigned scores closer to 1, while more casually offensive language is typically scored closer to the positive decision boundary of 0.5. Here, we consider two definitions for antisocial comments: if the comment's score is above the standard decision boundary (0.5) or a higher boundary (0.8) for highly toxic content.  As metrics, we include both (i) the count of comments exceeding both threshold as well as (ii) the percentage of non-toxic comments; the latter percentage allows us to model large discussions where some sub-discussions turn antisocial, but the majority of content is not antisocial.

We note that toxicity itself can be challenging to measure \cite{vidgen2019challenges}. Multiple models have been proposed for handling different aspects of antisocial behavior  \cite{fortuna2018survey} and that these models frequently have gaps in what they recognize \cite{arango2019hate,jurgens2019just}, can encode biases with respect to social groups \cite{sap2019risk}, and be susceptible to adversarial attacks \cite{grondahl2018all}. Nevertheless, as a widely-used measure, Perspective API provides a replicable---though imperfect---tool specifically designed for recognizing multiple forms of toxicity in online conversations in comments, which is the medium studied here.

\section{Problem Definition and Data}
\label{sec:data}

\begin{table}[!tb]
\centering

\begin{tabular}{r rrr}
 & \multicolumn{1}{c}{\textbf{Train}} & \multicolumn{1}{c}{\textbf{Dev}} & \multicolumn{1}{c}{\textbf{Test}} \\
\hline
\tlc & 4,290,361 & 10,844,404& 10,844,281\\
Subreddits & 11,992 & 53,675 & 53,650\\
\end{tabular}
\caption{Dataset sizes; note that training data has been downsampled for computational tractability. }
\label{tab:dataset}
\end{table}

This study's goal is to forecast the future behavior of a conversation from early signals. Specifically, we aim to predict whether the initial comment in a conversation will signal eventual prosocial behavior. This forecasting goal mirrors analogous work in antisocial behavior on conversational trajectories \citep{zhang2018conversations}, as a first step towards understanding prosocial evolution in conversations.\footnote{We also acknowledge that other setups have used more conversation as context for forecasting, e.g., \citet{chang2019trouble} and \citet{liu2018forecasting}. While these too are valid setups, we focus on the initial comment intentionally to see whether emerging prosocial conversations can be quickly identified and prioritized.}  We pursue this goal using data from Reddit, a large social media platform where users create posts and participate in threaded conversations relating to that post.  Critically, Reddit provides millions of conversations across different communities, known as subreddits, which span a variety of topics and interaction styles.

To analyze conversations, we extract all conversational threads under a Top-level Comment (\tlc) to a post; such comments are typically made in response to the post itself and serve as conversation starters for the rest of the community.
We filter these conversations by removing those where  the \tlc (i) has more than 3500 words, as manual inspection showed these were frequently spam posts, or (ii) has been deleted by the user or removed by a moderator. 
Additionally, Reddit includes a small number of bot accounts that interact in these conversations (e.g., replying with the number of ``ooofs'' a user has made); to avoid any confounding effect of these bots, we remove all threads containing a comment by a bot account, drawn from a known list of bots\footnote{\url{https://www.reddit.com/r/autowikibot/wiki/redditbots}}
and a manually-curated list based on inspection of high-frequency-posting accounts. 

The final dataset was constructed from a single year of Reddit activity using the $\sim$500M comments posted between January 2017 and December 2017. 
This process yielded 140.5M total conversations (unique \tlc) across 66K subreddits randomly partitioned into training, development, and test sets using an 8:1:1 ratio (Table \ref{tab:dataset}). Due to computational limitations in labeling each conversation with all metrics (e.g., scoring all comments using the rate-limited Perspective API), we downsample the training data to 4.3M conversations, requiring all subreddits to have at least 100 conversations and keeping at most 500 conversations per subreddit.

\section{Judging Conversation Prosociality}
\label{sec:prosocial-judgments}

What types of conversations do people find more prosocial? The multifaceted nature of prosociality makes numerically rating conversations challenging. Therefore, we answer this question by framing the rating as a paired choice task: given two conversations, select which conversation contains more prosocial behavior.

This binary rating setup also allows us to directly evaluate whether each proposed prosocial metric aligns with human judgments. 
For each metric (\sref{sec:prosocial}), we measure the strength of association with human judgments by computing Matthews Correlation Coefficient (MCC) from a 2x2 contingency table of which conversation in the pair had a higher metric score versus which was selected by annotators as being more prosocial.\footnote{While related to the $\chi^2$ measure, MCC differs in that it measure the \textit{strength} of association, rather than just whether the difference in the ratings is statistically significant.}

Annotated data was collected in two phases. An initial 2000 conversation pairs were collected by sampling two \tlc made to the same post using two strategies: (1) 1000 pairs of conversations made any time after the post was authored and (2) 1000 pairs made within 5 minutes of their post. 
After this initial selection, an additional 388 pairs of conversations were included to ensure that each metric occurred in at least 100 pairs, with the exception of the Donation metric, which only occurred in 78 pairs.
Two annotators participated in three rounds of training and then divided up the annotations. Annotators attained a Krippendorff's $\alpha$  of 0.78 on 300 mutually-labeled pairs, indicating high agreement.

\mypara{Results}
Table \ref{tab:metrics-table} summarizes all of the metrics and their correlations, revealing that most of the metrics predicted from theory are significantly associated with human judgments of prosociality.  
Four trends merit noting.
First, metrics for the breadth and depth of conversations were most correlated with prosocial judgments, with the strongest association for sustained conversation between people; these behavior promote social cohesion and, given the discussion-focused nature of Reddit, are easily measured in conversation.
Second, the second-most associated category was for information-providing behaviors.  While less common in Reddit conversations as a whole, these actions help other users meet their information needs.
Third, surprisingly, prior metrics suggested for prosocial behaviors, politeness and supportiveness, were positive but not significantly associated. We view this negative result as requiring further investigation to confirm, as more precise models for measuring these behaviors may provide more insight.
Fourth, the metric around toxicity showed that, indeed, the absence of toxicity was only moderately correlated with human judgments of prosocial conversations, while other types of behavior were more associated. Further, the presence of extremely toxic language, though rare, was not correlated, indicating that a broader picture of toxic behavior---not just extreme events---is necessary for prosocial judgment. This result also indicates that for platforms to measure their health, new metrics like those proposed above are needed and not simply measure the absence of antisocial behavior.  However, as our adopted measure of toxicity is a coarse-grain estimate, a potential avenue for future work is to examine whether the absence of specific forms of antisocial behavior or toxicity might individually be associated with the perception of prosocial behavior.

\mypara{Synthesizing Prosociality}
Prosocial behaviors often share a common motivation as individuals seek to engage constructively with one another. Thus, with their conceptual and thematic similarities, a single conversation may contain several of these prosocial behaviors. Given their shared motivation, we ask whether the prosocial metrics could be summarized with a single proxy metric? To test this, we adopt the approach of \cite{voigt2017language} for synthesizing a single metric from related prosocial behaviors around respect and compute the Principal Component Analysis (PCA) over all the metric scores for conversations in the training data. PCA decomposes these in a set of underlying latent behaviors, capturing the inherent correlations between metrics.

\begin{figure}[t]
    \centering
    \includegraphics[width=0.48\textwidth]{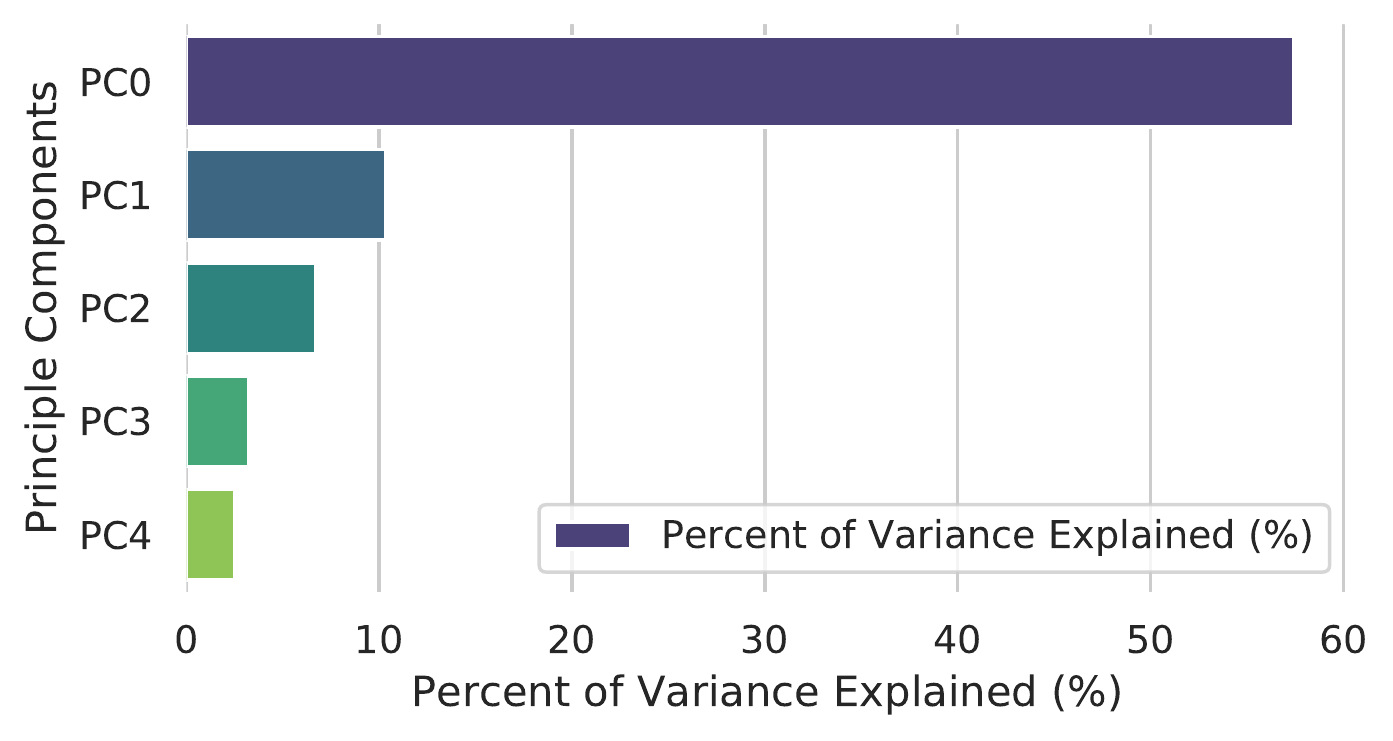}
    \caption{The percentage of variance explained by the first five principal components across the values of prosocial metrics for conversations shows that the first PCA component explains 57.4\% of the variance in the prosocial metrics' values, indicating many prosocial metrics are highly correlated. The loadings for the first and second components are shown in Appendices Figures \ref{fig:loadingforpc0} and \ref{fig:loadingforpc1}.}
    \label{fig:explained-variance-from-pca}
\end{figure}

As shown in Figure \ref{fig:explained-variance-from-pca}, the first PCA component explains 57.4\% of the variance in the prosocial metrics' values and positively loads on all prosocial metrics (Appendix Figure \ref{fig:loadingforpc0}), suggesting the component effectively captures shared behavior. In contrast, the second principal component captures roughly 10\% of the variance, with its loading not reflective of a coherent set of prosocial behaviors (Appendix Figure \ref{fig:loadingforpc1}). Thus, while a simplification of the inherent complexity of online behavior, the first principal component offers a compelling single value to act as a proxy in comparing the variety of behaviors seen in conversations. In this sense, the component acts analogously to other high-level estimates of behavior such as the toxicity score measured by Perspective API \cite{wulczyn2017ex} that provides a single value for downstream applications.

To further test and validate the use of the first principal component as a proxy in our experiments, we calculate its MCC score with human judgments of prosociality, as was done for the individual metrics (Table \ref{tab:metrics-table}). The resulting MCC of 0.63 is higher than the MCC for any single metric and indicates the component's value is strongly reflective of human judgments of prosociality. Thus, given the single component that explains a substantial portion of the variance and its close correlation with human judgments, we view this first component as an effective single proxy for evaluating conversations. However, we recognize that the metrics studied here, while diverse, do not capture all of the prosociality, nor does the first component capture all of the variance, so that while strongly correlated, this first component is only an initial step at estimating prosociality.

\section{Forecasting Prosocial Behavior}
\label{sec:forecasting}

Given that the proposed metrics reflect human judgments of prosocial behavior,  we introduce computational models to forecast the degree of prosociality a conversation will ultimately from its initial discussion. 
Our ultimate motivation is to train a forecasting model on conversational outcomes using massive amounts of data labeled with computational-estimated prosociality and then, in \Sref{sec:ranking}, fine-tune this model to rank conversations based on human judgments.

\subsection{Task and Experimental Setup}


The first principal component strongly correlates with human judgments (\Sref{sec:prosocial-judgments}) and therefore we treat the first component's value  as a single numeric estimate of the prosociality of a conversation following a \tlc. We refer to this value as the \tlc's \textit{prosocial trajectory}. While using a single value to capture prosociality across all comments under a \tlc likely simplifies some nuances of different behaviors, the single value nonetheless provides a useful proxy for conversational quality akin to other antisocial metrics; further, the PCA analysis showed that a single component captured the majority of the variance, with no other component having a consistent or substantial loading on prosociality, suggesting that a single metric, while simplifying, could still be effective at reflecting broad trends in conversational prosociality.

Models are trained to predict the prosocial trajectory value given (i) the title and text of a post, (ii) the \tlc, and (iii) metadata for the comment including the subreddit and time when it was posted.
Models are fit and tested using the training, development, and test partitions shown in Table \ref{tab:dataset} using MSE as the objective.

\subsection{Features}
\label{sec:features}

Models are trained using four categories of features. The first includes features from all the prosocial metrics in \Sref{sec:prosocial}, with the exception of accommodation; these features provide some estimate of whether the conversation is beginning on a positive note.
The second category includes features reflecting the comment's relationship to the post: i) topic distributions of the post and \tlc, ii) cosine similarity of the two topic distributions, and iii) Jaccard similarity of non-stop word content in post and \tlc. For topics, a 20-topic LDA model was trained for post and \tlc text each using Mallet \cite{McCallumMALLET}.
The third category includes features of the \tlc: i) number of words, ii) sentiment iii) subjectivity, iv) number of misspelled words, v) Flesch-Kincaid reading score, and vi) author gender.
Finally, the fourth category reflects the circumstances in which the \tlc was made: i) the subreddit containing the \tlc ii) time features for the day of a month, day of a week, and hour of a day, and iii) minutes between the post's creation time and \tlc's creation time.

\subsection{Models}
\label{sec:models}


\mypara{Baselines}
As a first baseline, we train a linear regression with L2 loss over all the features in \Sref{sec:features}, using unigram and bigram  features for the \tlc text.
The second baseline uses XGBoost \cite{chen2016xgboost}, which allows us to test for combinatorial interactions between features. XGBoost was trained with a tree-based booster that had a learning rate $\eta$ of 0.05, L2 regularization $\lambda$ of 3.0, and L1 regularization $\alpha$ of 1.0. The minimum loss reduction $\gamma$ required to make a further partition on a leaf node of the tree was set to 1. The maximum depth of a tree was 4. The subsample ratio of the training instances and that of columns when constructing each tree were both 0.8. We trained the model for 5000 iterations, with one parallel tree constructed during each iteration.

\begin{figure}[t]
    \centering
    \includegraphics[width=0.48\textwidth]{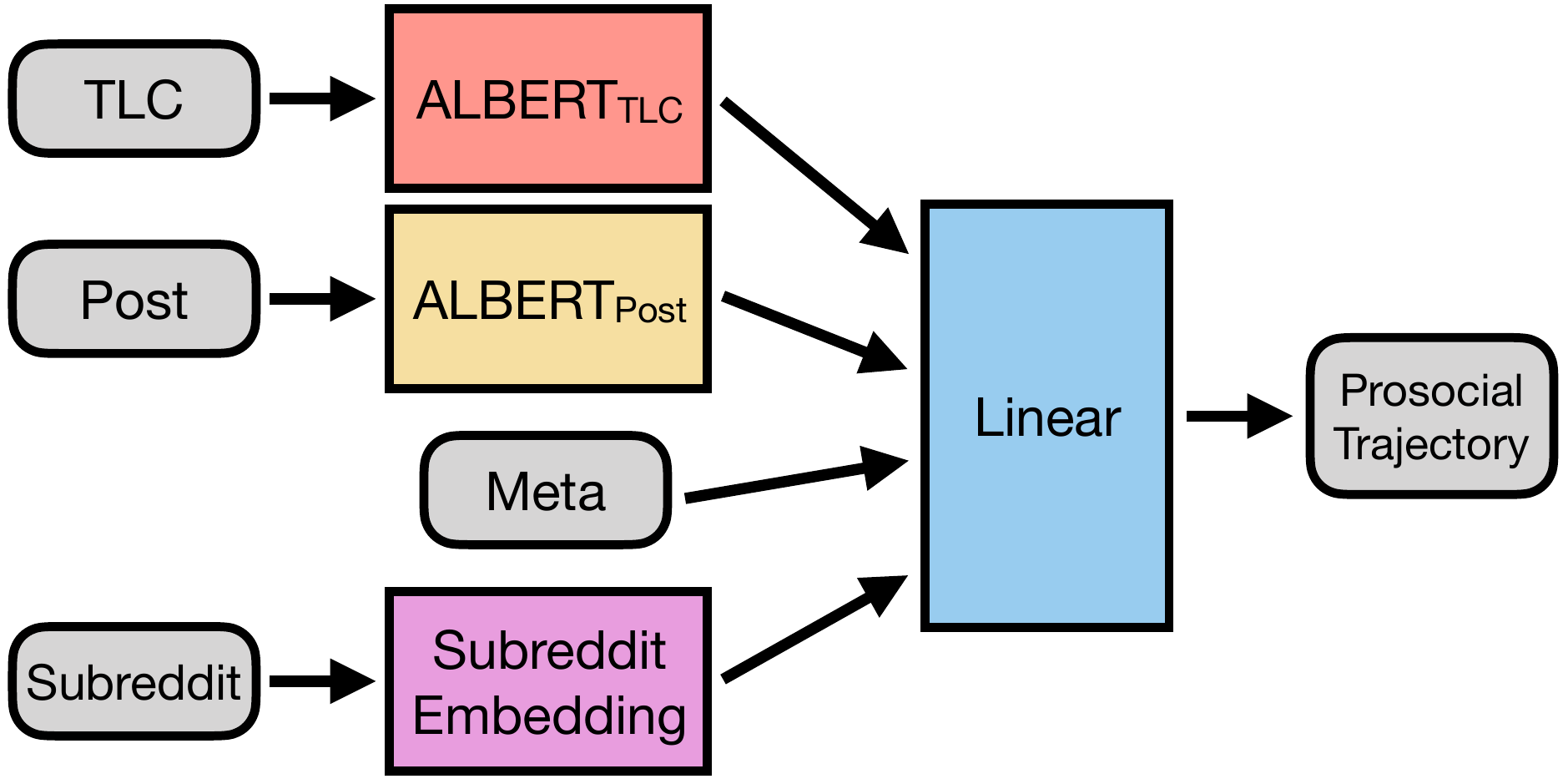}
    \caption{The proposed model for predicting a conversation's prosocial trajectory.}
    \label{fig:post-and-comment-model}
\end{figure}

\mypara{Our Models}
We introduce two trajectory-forecasting models built on top of the Albert model \cite{lan2019albert}, which is a refinement of the BERT pretrained language model \cite{devlin2018bert}.
In this model shown in \fref{fig:post-and-comment-model}, the post and \tlc are fed into separate Albert-based networks and the \verb|[CLS]| tokens from each are used as representations of their text. This vector is concatenated with a vector containing the non-textual features from \Sref{sec:features} to represent the entire input. The output layer consists of a linear layer. 
The subreddit is represented as an embedding; these embeddings are initialized from the 300-dimensional embeddings from \citet{kumar2019predicting} but reduced to 16 dimensions using PCA, which accounted for 72\% of the variance. A total of 5278 of our 11,993 subreddits had predefined embeddings from this process, with the remaining using random initialization.
To measure the effect of pre-training, we include a version of the model using only the off-the-shelf Albert parameters that are left unchanged and a second version that is first fine-tuned on Reddit post and \tlc text (respectively) using masked language modeling and then its parameters are updated during trajectory training.
Additional details on hyperparameters and training are in Appendix \ref{app:hyperparameters}.

\begin{figure}
    \centering
    \includegraphics[width=0.9\linewidth]{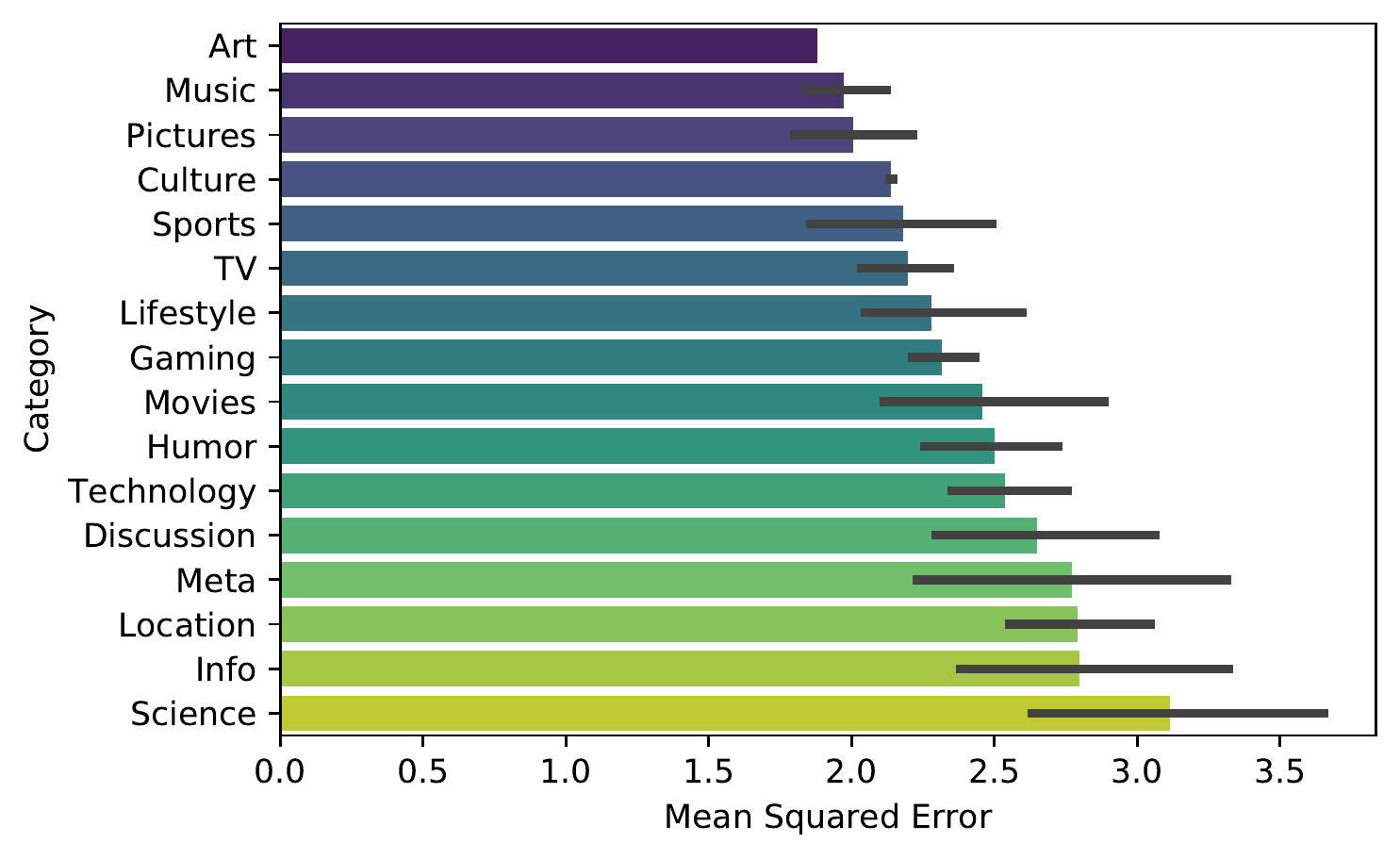}
    \caption{The MSE of prosocial forecasts within different subreddit categories shows  that our Albert model attains higher performance in communities whose discussion relates to pop-culture such as Movies, Art, and Culture. Mean scores for our model and XGBoost are reported in Table \ref{tab:category-mse-loss}.}
    \label{fig:category-mse-loss}
\end{figure}

\subsection{Results}

Models were able to identify sufficient signals of the conversational trajectory from just the post and \tlc to forecast its eventual value, as shown in Table \ref{tab:results}. While performance is low, high performance is not expected in this setting as models only have access to the start of a conversation, which can take many potential trajectories. Nevertheless, the substantial improvement of the XGBoost and our full model over both the mean-value and linear regression baselines indicate that some signals can be reliably found which would aid in proactive conversation sorting.
Examining relative differences between the deep learning models, fine-tuning the language model parameters was critical to performance improvement, with the baseline Linear Regression outperforming the substantially more complicated model that used off-the-shelf parameters. 

Conversations in some topics may be easier to forecast than others. To test for topical effects, we use the subreddit categorization from \url{http://redditlist.com/} and compute within-category MSE. A clear trend emerges where both the XGBoost and our model performed better than average for categories related to pop-culture such as Movies, Art, and Culture (p$<$0.01 using  the Kolmogorov-Smirnov test on error distributions). Figure \ref{fig:category-mse-loss} shows the mean MSE per \tlc within each  category for our model, with Appendix Table \ref{tab:category-mse-loss} reporting means for both models.  In contrast, both models performed worse for Science and Information subreddits which may take on different discussion patterns. These results highlight that different communities each have their own norms, which can make estimating conversational trajectory easier. 

For our model, the ten subreddits with the lowest MSE included \textit{r/aww, r/funny}, and \textit{r/NatureIsFuckingLit}, three highly popular subreddits with millions of subscribers, suggesting the model performs well in lighthearted discussions. In contrast, the ten subreddits with the highest MSE included, \textit{r/changemyview, r/PoliticalDiscussion}, and \textit{r/geopolitics}, three popular subreddits that feature long, often-contentious discussions. We speculate that conversations in those communities are more unpredictable due to the inherent tension around the topics and therefore reflect a significant challenge to forecasting models.

\begin{figure}
    \centering
    \includegraphics[width=0.9\linewidth]{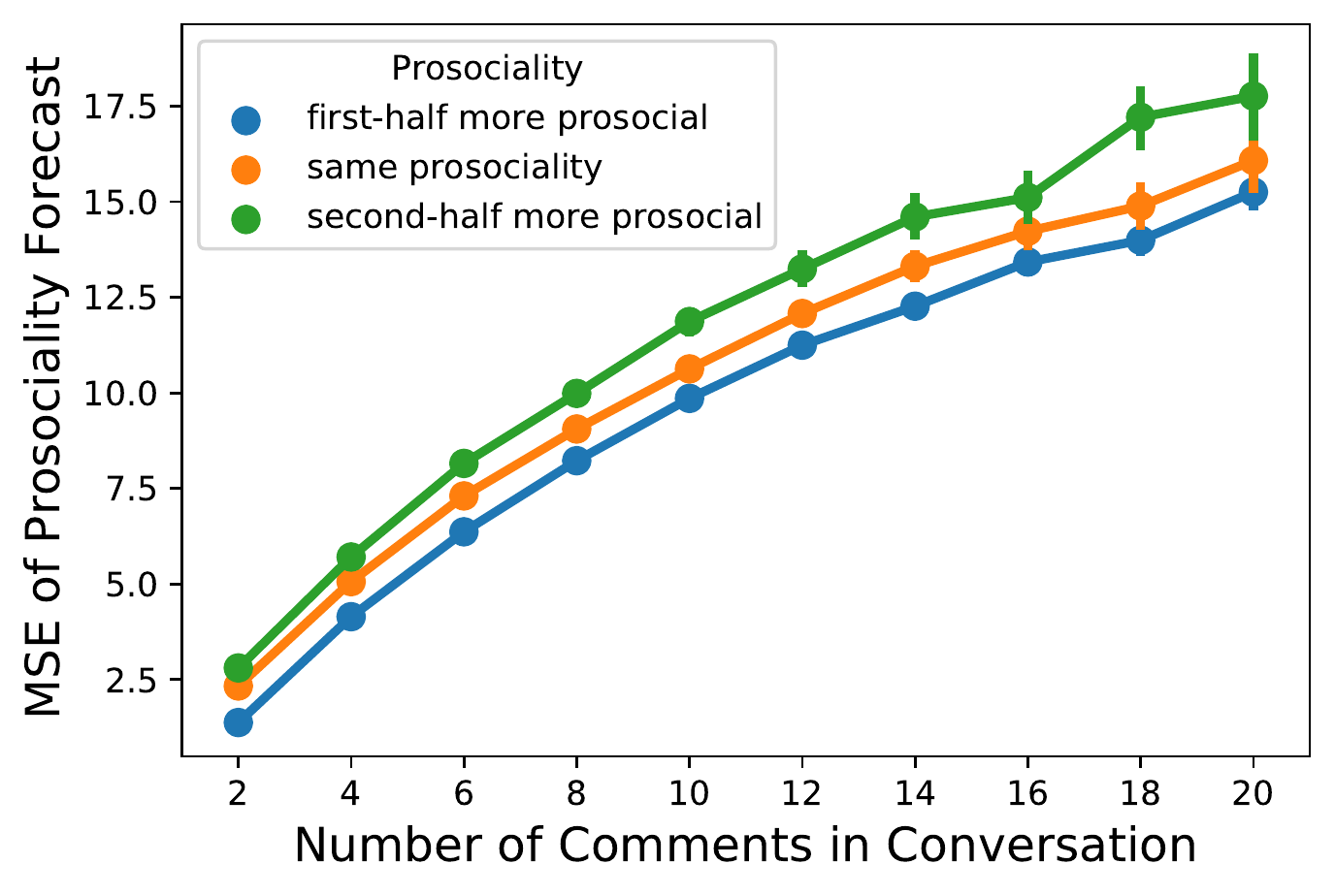}
    \caption{Error in forecasting prosociality relative to when the prosocial occurred within the subsequent conversation.}
    \label{fig:fh-sh-error}
\end{figure}

Prosocial behavior may occur at any point in the subsequent conversation, which creates a challenge for our model that forecasts from only the initial comment. As a follow-up analysis, we test  how our model's error changes relative to when the prosocial behavior occurs. We sampled 307K conversations of even length $n$ and measure the prosociality (via the first principal  component) of the first $\frac{n}{2}$ comments and last $\frac{n}{2}$  comments, ordered temporally. We then stratify the conversations relative to whether the first half was more prosocial, less prosocial, or even. \fref{fig:fh-sh-error} reveals that across conversation sizes (shown up to 20 comments), our model consistently has lower error for prosocial behavior that occurs soon after the initial \tlc. Across all  sizes, conversations with early prosociality have lower error (mean MSE 5.96) than those with later prosociality (mean MSE 7.81; p$<$0.01 using Wilcoxon), suggesting that models that use increasing amounts of context beyond just a \tlc \citep[e.g,][]{chang2019trouble,liu2018forecasting} would perform well. $\sim$8\% of the conversations had the same estimated prosociality in each half, which suggests that future work could identify new dimensions or refine the tools of our existing measures to better discriminate between such cases.

\begin{table}[tb]
  \centering
    \begin{tabular}{r cc }
     Model & MSE & R$^2$ \\
     \hline
     \textit{Mean Value} (baseline)          & 3.010 & -0.003 \\
     Linear Regression              & 2.393 & 0.209 \\
     XGBoost                        & 2.209 & 0.269 \\ 
     Our model (frozen Albert)      & 2.209 & 0.157 \\
     Our model  & 2.230 & 0.262 \\ 
    \end{tabular}
  \caption{Mean squared Error and R$^2$ for forecasting prosocial conversational trajectory shows that models can estimate the   trajectory value from early signals.}
  \label{tab:results}%
\end{table}%

\section{Ranking Conversations}
\label{sec:ranking}

We have shown that the prosocial trajectory of a conversation can be forecasted from the linguistic and social signals in its first comment. Here, we test whether these models can be used to rank conversations by their potential outcome. We adopt a simplified ranking setting where a model (or human) is shown a post and two of its \tlc and asked to select which \tlc is likely to lead to a more prosocial conversation.

\subsection{Data}
\label{sec:human-preds}

Data was drawn from the 2388 instances annotated in \Sref{sec:prosocial-judgments} for which conversation was more prosocial.\footnote{Note that during the previous annotation, the starts of these conversations were never shown to annotators.}
An additional 1000 pairs were annotated where one of the \tlc received no replies.\footnote{While empty conversations would seem to always be less prosocial, annotators preferred such conversations preferred to those containing mostly toxic comments, though such cases were rare in practice.} 
This data was partitioned into 80\% training, 10\% development, and 10\% test, stratifying across the three sampling strategies used to create it.
Two annotators rated pairs of \tlc to the same post according to their judgments of which were more likely to result in a positive, prosocial outcome.  
Annotators had an Krippendorff's $\alpha$=0.59. As the task is inherently difficult with no objective ground truth present in the \tlc, high agreement is not expected; however, the moderate agreement indicates that annotators were able to consistently identify a common set of  lexical features they considered predictive.

\subsection{Models}

\begin{figure}
    \centering
    \includegraphics[width=0.9\linewidth]{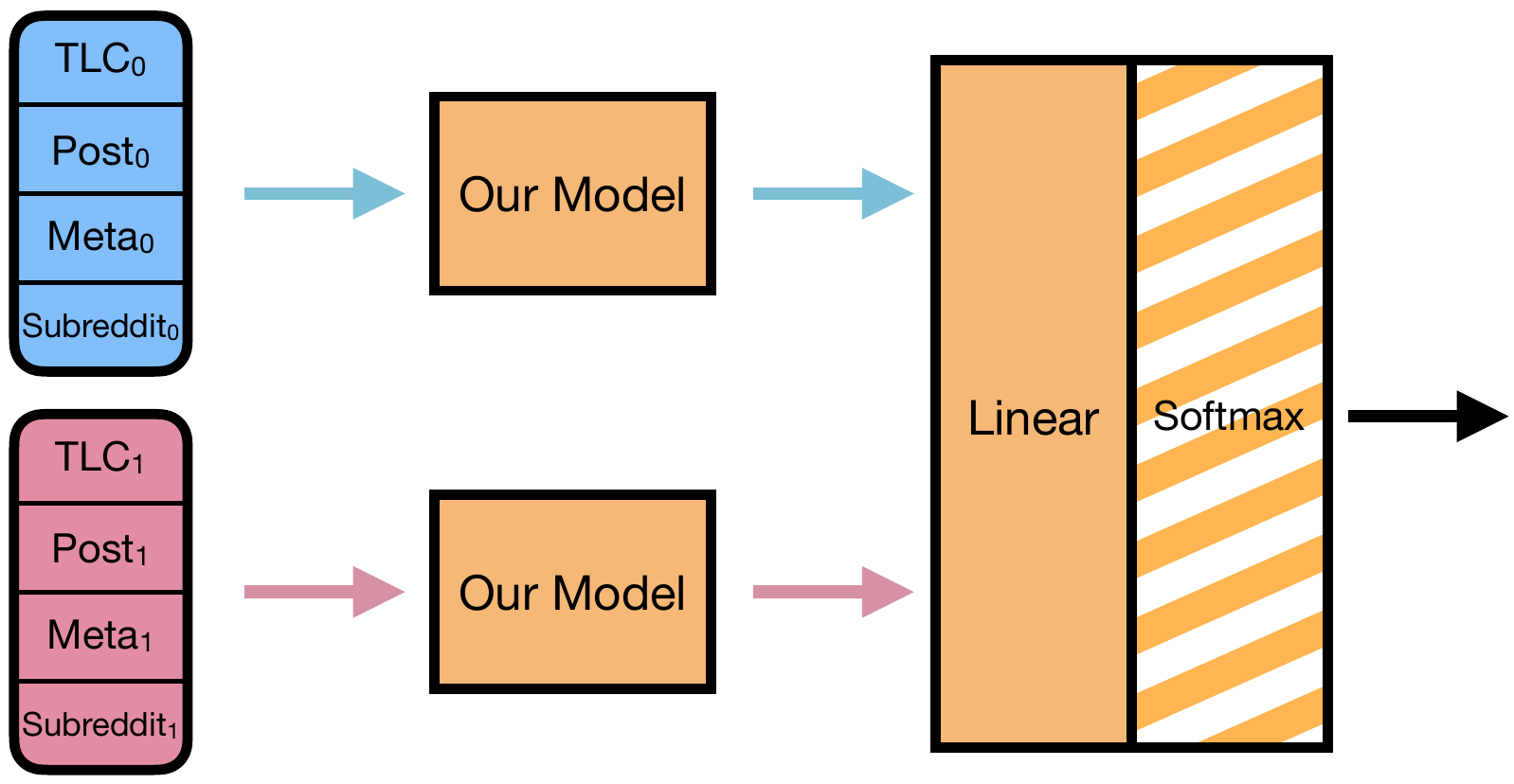}
    \caption{Diagram of our deep learning model for forecasting which conversation will be more prosocial. }
    \label{fig:twhich-is-more-model}
\end{figure}

Our proposed ranking model, shown in \fref{fig:twhich-is-more-model}, fuses two of the post-and-comment forecasting models (\fref{fig:post-and-comment-model}) with a linear layer to allow fine-tuning on held out human-judgments. 
To measure the effect of pre-training, we include a model that directly uses the trajectory estimates of the component models and selects the \tlc with a higher estimate; a similar model is included for XGBoost.
As a baseline comparison, we include a logistic regression classifier trained on unigram and bigrams directly on the \tlc.
Finally, we include an \textit{oracle} classifier that uses the actual trajectory value for each \tlc and selects the higher-valued; this oracle-based classifier reflects the upper bound for performance if forecasting models would perfectly estimate trajectory and rank using only that value.

\subsection{Results}

Models were able to surpass human performance at correctly selecting the conversation that would result in more prosocial behavior, as shown in Table \ref{tab:ranking-results}. Consistent with prior work on predicting antisocial behavior in early conversations, high performance is not expected in this tightly-constrained setting  \cite{liu2018forecasting,zhang2018conversations,jiao2018find}, as new people join conversations each with their own interests and motivations that affect the trajectory. However, the moderate performance on this difficult task suggests that models can reliably pick up on prosocial signals from the very first comment in a discussion, which is sufficient for aiding in re-ranking newly-started conversations.

The pre-trained forecasting model's accuracy is the primary driver for performance. 
We can observe a soft upper bound for performance by comparing the model with the ranking prediction performance of a model that perfectly predicts the trajectory, shown as the \textit{oracle trajectory prediction}. If a forecasting model would be able to forecast the trajectory with perfect accuracy (like this oracle), simply picking the conversation with the higher estimated trajectory would achieve an 86.5\% accuracy at selecting the conversation that ultimately had more prosocial behavior. Although exact trajectory estimates are unlikely from the \tlc alone, this illustration's result suggests that higher ranking performance is possible in future models, such as those with more data from incrementally forecasting as the conversation grows after its initial stages; indeed, models for forecasting antisocial behavior  from longer context suggest that such a result is possible \cite{chang2019trouble}.

Despite having moderate agreement on which \tlc was predicted to have a more prosocial outcome, humans performed worse than the proposed model. Annotators and the best model had only weak agreement in their judgments ($\alpha$=0.29) and with 69.7\% of the annotator's correct decisions also being selected by the model. This result, combined with the inter-annotator agreement, suggests that annotators were able to pick up on a complementary set of linguistic signals not used by the model, which future models might identify to improve performance.

\begin{table}[tb]
  \centering
    \begin{tabular}{r c  }
     \textbf{Model} & \textbf{Accuracy}  \\
     \hline
     Logistic Regression on \tlc & 0.463  \\
     XGBoost &  0.540  \\ 
     \textit{Human Prediction} & 0.563 \\
     Our model (only pre-trained) & 0.566  \\ 
     Our model (fine-tuned model)  & \textbf{0.578}  \\ 
     \textit{Oracle Trajectory Prediction} & 0.865  \\
    \end{tabular}%
  \caption{Performances at predicting which of two conversations will have a more prosocial outcome shows that our best computation model  outperforms human predictions.}
  \label{tab:ranking-results}%
\end{table}%

\section{Discussion}

Prosociality can take many forms and in this paper, we have developed classifiers to recognize a variety of these behaviors, showing they can be recognized and that many are correlated with each other. However, there are multiple directions that could be taken to better reflect prosociality as a whole. First, our model is agnostic to the  community itself in considering what behavior should be considered prosocial, even though communities are known to have different social norms \cite{chandrasekharan2018internet,chancellor2018norms,matias2019preventing}; for example, the jocular nature of sports and gaming communities may consider politeness out of the ordinary and not inline with their desired prosocial behaviors. This direction is further supported by the observed variance in performance across different categories of subreddits (Figure \ref{fig:category-mse-loss}), which suggests that  directly incorporating the norms of specific communities could improve performance.
Second, while our PCA score unifies many prosocial behaviors under a single metric (much like general ``toxicity'' scores), a significant amount of variation remains unexplained. The PCA value used in this paper offers a compelling and practical operationalization. However, further analyses are needed to identify other prototypical forms of prosociality and their effect on conversations. 
Third, our forecast and ranking models simplify the task to only looking at a \tlc in predicting  conversational trajectory. As conversations can often take unpredictable turns, these later comments are likely to influence its prosocial trajectory, which cannot be observed from the \tlc alone. However, given our promising results on just the \tlc, later models may improve upon these results through iteratively predicting prosocial trajectory from the growing sequence of comments in a conversation, as others have done in forecasting antisocial behavior \cite{liu2018forecasting,chang2019trouble}.

\section{Conclusion}

Online conversations can take many trajectories, not all of them pleasant. Improving our ability to recognize and highlight nascent \textit{prosocial} conversations early on in their discussion can directly impact the daily lives and discussions of millions by fostering a more amicable and productive discourse online. This paper has introduced a series of metrics for different forms of prosocial behavior and accompanying computational techniques for recognizing them, showing that these behaviors are strongly correlated with human judgments of prosocial conversations---and that prosociality isn't simply explained by the absence of antisocial behavior.  In two experiments, we introduce a series of deep learning models showing that prosocial trajectories can be forecasted from just the initial comment in a conversation. Then, we show that these models can be adapted to predict which of two conversations is more likely to have a prosocial outcome from these signals, providing a ranking mechanism for increasing the visibility of conversations likely to have prosocial outcomes. While forecasting from such little data is difficult---but critical to ranking new conversations---our model is able to substantially improve on human performance over chance (24\%) at selecting the one with better outcome. Our model provides an initial starting point for conversation ranking and we show that if the forecast was completely accurate, such models would have an upper limit of 86.5\%, further motivating work in this area. Code, data, and models are available at \url{https://github.com/davidjurgens/prosocial-conversation-forecasting}.

\begin{acks}
This material is based upon work supported by the National Science Foundation under Grants No 1850221 and 2007251.
\end{acks}

\bibliographystyle{ACM-Reference-Format}
\bibliography{references}


 \newcommand{\noop}[1]{}
\begin{thebibliography}{86}


\ifx \showCODEN    \undefined \def \showCODEN     #1{\unskip}     \fi
\ifx \showDOI      \undefined \def \showDOI       #1{#1}\fi
\ifx \showISBNx    \undefined \def \showISBNx     #1{\unskip}     \fi
\ifx \showISBNxiii \undefined \def \showISBNxiii  #1{\unskip}     \fi
\ifx \showISSN     \undefined \def \showISSN      #1{\unskip}     \fi
\ifx \showLCCN     \undefined \def \showLCCN      #1{\unskip}     \fi
\ifx \shownote     \undefined \def \shownote      #1{#1}          \fi
\ifx \showarticletitle \undefined \def \showarticletitle #1{#1}   \fi
\ifx \showURL      \undefined \def \showURL       {\relax}        \fi
\providecommand\bibfield[2]{#2}
\providecommand\bibinfo[2]{#2}
\providecommand\natexlab[1]{#1}
\providecommand\showeprint[2][]{arXiv:#2}

\bibitem[\protect\citeauthoryear{Akerlof and Kranton}{Akerlof and
  Kranton}{2000}]%
        {akerlof2000economics}
\bibfield{author}{\bibinfo{person}{George~A Akerlof} {and}
  \bibinfo{person}{Rachel~E Kranton}.} \bibinfo{year}{2000}\natexlab{}.
\newblock \showarticletitle{Economics and identity}.
\newblock \bibinfo{journal}{\emph{The Quarterly Journal of Economics}}
  \bibinfo{volume}{115}, \bibinfo{number}{3} (\bibinfo{year}{2000}),
  \bibinfo{pages}{715--753}.
\newblock


\bibitem[\protect\citeauthoryear{Algoe}{Algoe}{2012}]%
        {algoe2012find}
\bibfield{author}{\bibinfo{person}{Sara~B Algoe}.}
  \bibinfo{year}{2012}\natexlab{}.
\newblock \showarticletitle{Find, remind, and bind: The functions of gratitude
  in everyday relationships}.
\newblock \bibinfo{journal}{\emph{Social and Personality Psychology Compass}}
  \bibinfo{volume}{6}, \bibinfo{number}{6} (\bibinfo{year}{2012}),
  \bibinfo{pages}{455--469}.
\newblock


\bibitem[\protect\citeauthoryear{Arango, P{\'e}rez, and Poblete}{Arango
  et~al\mbox{.}}{2019}]%
        {arango2019hate}
\bibfield{author}{\bibinfo{person}{Aym{\'e} Arango}, \bibinfo{person}{Jorge
  P{\'e}rez}, {and} \bibinfo{person}{Barbara Poblete}.}
  \bibinfo{year}{2019}\natexlab{}.
\newblock \showarticletitle{Hate speech detection is not as easy as you may
  think: A closer look at model validation}. In
  \bibinfo{booktitle}{\emph{Proceedings of SIGIR}}. \bibinfo{pages}{45--54}.
\newblock


\bibitem[\protect\citeauthoryear{Bachorowski and Owren}{Bachorowski and
  Owren}{2001}]%
        {bachorowski2001not}
\bibfield{author}{\bibinfo{person}{Jo-Anne Bachorowski} {and}
  \bibinfo{person}{Michael~J Owren}.} \bibinfo{year}{2001}\natexlab{}.
\newblock \showarticletitle{Not all laughs are alike: Voiced but not unvoiced
  laughter readily elicits positive affect}.
\newblock \bibinfo{journal}{\emph{Psychological Science}} \bibinfo{volume}{12},
  \bibinfo{number}{3} (\bibinfo{year}{2001}), \bibinfo{pages}{252--257}.
\newblock


\bibitem[\protect\citeauthoryear{Bak, Lin, and Oh}{Bak et~al\mbox{.}}{2014}]%
        {bak2014self}
\bibfield{author}{\bibinfo{person}{JinYeong Bak}, \bibinfo{person}{Chin-Yew
  Lin}, {and} \bibinfo{person}{Alice Oh}.} \bibinfo{year}{2014}\natexlab{}.
\newblock \showarticletitle{Self-disclosure topic model for classifying and
  analyzing Twitter conversations}. In \bibinfo{booktitle}{\emph{Proceedings of
  EMNLP}}. \bibinfo{pages}{1986--1996}.
\newblock


\bibitem[\protect\citeauthoryear{Bar-Tal}{Bar-Tal}{1976}]%
        {bar1976prosocial}
\bibfield{author}{\bibinfo{person}{Daniel Bar-Tal}.}
  \bibinfo{year}{1976}\natexlab{}.
\newblock \bibinfo{booktitle}{\emph{Prosocial behavior: Theory and research.}}
\newblock \bibinfo{publisher}{Hemisphere Publishing Corp}.
\newblock


\bibitem[\protect\citeauthoryear{Barak and Gluck-Ofri}{Barak and
  Gluck-Ofri}{2007}]%
        {barak2007degree}
\bibfield{author}{\bibinfo{person}{Azy Barak} {and} \bibinfo{person}{Orit
  Gluck-Ofri}.} \bibinfo{year}{2007}\natexlab{}.
\newblock \showarticletitle{Degree and reciprocity of self-disclosure in online
  forums}.
\newblock \bibinfo{journal}{\emph{CyberPsychology \& Behavior}}
  \bibinfo{volume}{10}, \bibinfo{number}{3} (\bibinfo{year}{2007}),
  \bibinfo{pages}{407--417}.
\newblock


\bibitem[\protect\citeauthoryear{Bartlett and DeSteno}{Bartlett and
  DeSteno}{2006}]%
        {bartlett2006gratitude}
\bibfield{author}{\bibinfo{person}{Monica~Y Bartlett} {and}
  \bibinfo{person}{David DeSteno}.} \bibinfo{year}{2006}\natexlab{}.
\newblock \showarticletitle{Gratitude and prosocial behavior: Helping when it
  costs you}.
\newblock \bibinfo{journal}{\emph{Psychological science}} \bibinfo{volume}{17},
  \bibinfo{number}{4} (\bibinfo{year}{2006}), \bibinfo{pages}{319--325}.
\newblock


\bibitem[\protect\citeauthoryear{Batson and Powell}{Batson and Powell}{2003}]%
        {batson2003altruism}
\bibfield{author}{\bibinfo{person}{C~Daniel Batson} {and}
  \bibinfo{person}{Adam~A Powell}.} \bibinfo{year}{2003}\natexlab{}.
\newblock \showarticletitle{Altruism and prosocial behavior}.
\newblock \bibinfo{journal}{\emph{Handbook of psychology}}
  (\bibinfo{year}{2003}), \bibinfo{pages}{463--484}.
\newblock


\bibitem[\protect\citeauthoryear{Berry}{Berry}{2009}]%
        {berry2009you}
\bibfield{author}{\bibinfo{person}{Roger Berry}.}
  \bibinfo{year}{2009}\natexlab{}.
\newblock \showarticletitle{You could say that: the generic second-person
  pronoun in modern English}.
\newblock \bibinfo{journal}{\emph{English Today}} \bibinfo{volume}{25},
  \bibinfo{number}{3} (\bibinfo{year}{2009}), \bibinfo{pages}{29--34}.
\newblock


\bibitem[\protect\citeauthoryear{Biyani, Caragea, Mitra, and Yen}{Biyani
  et~al\mbox{.}}{2014}]%
        {biyani2014identifying}
\bibfield{author}{\bibinfo{person}{Prakhar Biyani}, \bibinfo{person}{Cornelia
  Caragea}, \bibinfo{person}{Prasenjit Mitra}, {and} \bibinfo{person}{John
  Yen}.} \bibinfo{year}{2014}\natexlab{}.
\newblock \showarticletitle{Identifying emotional and informational support in
  online health communities}. In \bibinfo{booktitle}{\emph{Proceedings of
  COLING}}.
\newblock


\bibitem[\protect\citeauthoryear{Brown and Smart}{Brown and Smart}{1991}]%
        {brown1991self}
\bibfield{author}{\bibinfo{person}{Jonathon~D Brown} {and} \bibinfo{person}{S
  Smart}.} \bibinfo{year}{1991}\natexlab{}.
\newblock \showarticletitle{The self and social conduct: Linking
  self-representations to prosocial behavior.}
\newblock \bibinfo{journal}{\emph{Journal of Personality and Social
  psychology}} \bibinfo{volume}{60}, \bibinfo{number}{3}
  (\bibinfo{year}{1991}), \bibinfo{pages}{368}.
\newblock


\bibitem[\protect\citeauthoryear{Brown}{Brown}{2015}]%
        {brown2015politeness}
\bibfield{author}{\bibinfo{person}{Penelope Brown}.}
  \bibinfo{year}{2015}\natexlab{}.
\newblock \showarticletitle{Politeness and language}.
\newblock In \bibinfo{booktitle}{\emph{The International Encyclopedia of the
  Social and Behavioural Sciences (IESBS),(2nd ed.)}}.
  \bibinfo{publisher}{Elsevier}, \bibinfo{pages}{326--330}.
\newblock


\bibitem[\protect\citeauthoryear{Bucher}{Bucher}{2012}]%
        {bucher2012want}
\bibfield{author}{\bibinfo{person}{Taina Bucher}.}
  \bibinfo{year}{2012}\natexlab{}.
\newblock \showarticletitle{Want to be on the top? Algorithmic power and the
  threat of invisibility on Facebook}.
\newblock \bibinfo{journal}{\emph{New media \& society}} \bibinfo{volume}{14},
  \bibinfo{number}{7} (\bibinfo{year}{2012}), \bibinfo{pages}{1164--1180}.
\newblock


\bibitem[\protect\citeauthoryear{Buechel, Buffone, Slaff, Ungar, and
  Sedoc}{Buechel et~al\mbox{.}}{2018}]%
        {buechel2018modeling}
\bibfield{author}{\bibinfo{person}{Sven Buechel}, \bibinfo{person}{Anneke
  Buffone}, \bibinfo{person}{Barry Slaff}, \bibinfo{person}{Lyle Ungar}, {and}
  \bibinfo{person}{Joao Sedoc}.} \bibinfo{year}{2018}\natexlab{}.
\newblock \showarticletitle{Modeling empathy and distress in reaction to news
  stories}. In \bibinfo{booktitle}{\emph{Proceedings of IJCNLP}}.
\newblock


\bibitem[\protect\citeauthoryear{Burrow and Rainone}{Burrow and
  Rainone}{2017}]%
        {burrow2017many}
\bibfield{author}{\bibinfo{person}{Anthony~L Burrow} {and}
  \bibinfo{person}{Nicolette Rainone}.} \bibinfo{year}{2017}\natexlab{}.
\newblock \showarticletitle{How many likes did I get?: Purpose moderates links
  between positive social media feedback and self-esteem.}
\newblock \bibinfo{journal}{\emph{Journal of Experimental Social Psychology}}
  \bibinfo{volume}{69} (\bibinfo{year}{2017}), \bibinfo{pages}{232--236}.
\newblock


\bibitem[\protect\citeauthoryear{Chancellor, Hu, and De~Choudhury}{Chancellor
  et~al\mbox{.}}{2018}]%
        {chancellor2018norms}
\bibfield{author}{\bibinfo{person}{Stevie Chancellor}, \bibinfo{person}{Andrea
  Hu}, {and} \bibinfo{person}{Munmun De~Choudhury}.}
  \bibinfo{year}{2018}\natexlab{}.
\newblock \showarticletitle{Norms matter: contrasting social support around
  behavior change in online weight loss communities}. In
  \bibinfo{booktitle}{\emph{Proceedings of CHI}}. \bibinfo{pages}{1--14}.
\newblock


\bibitem[\protect\citeauthoryear{Chandrasekharan, Samory, Jhaver, Charvat,
  Bruckman, Lampe, Eisenstein, and Gilbert}{Chandrasekharan
  et~al\mbox{.}}{2018}]%
        {chandrasekharan2018internet}
\bibfield{author}{\bibinfo{person}{Eshwar Chandrasekharan},
  \bibinfo{person}{Mattia Samory}, \bibinfo{person}{Shagun Jhaver},
  \bibinfo{person}{Hunter Charvat}, \bibinfo{person}{Amy Bruckman},
  \bibinfo{person}{Cliff Lampe}, \bibinfo{person}{Jacob Eisenstein}, {and}
  \bibinfo{person}{Eric Gilbert}.} \bibinfo{year}{2018}\natexlab{}.
\newblock \showarticletitle{The Internet's Hidden Rules: An Empirical Study of
  Reddit Norm Violations at Micro, Meso, and Macro Scales}.
\newblock \bibinfo{journal}{\emph{Proceedings of CSCW}} (\bibinfo{year}{2018}).
\newblock


\bibitem[\protect\citeauthoryear{Chang and Danescu-Niculescu-Mizil}{Chang and
  Danescu-Niculescu-Mizil}{2019}]%
        {chang2019trouble}
\bibfield{author}{\bibinfo{person}{Jonathan~P Chang} {and}
  \bibinfo{person}{Cristian Danescu-Niculescu-Mizil}.}
  \bibinfo{year}{2019}\natexlab{}.
\newblock \showarticletitle{Trouble on the horizon: Forecasting the derailment
  of online conversations as they develop}. In
  \bibinfo{booktitle}{\emph{Proceedings of EMNLP}}.
\newblock


\bibitem[\protect\citeauthoryear{Chen and Guestrin}{Chen and Guestrin}{2016}]%
        {chen2016xgboost}
\bibfield{author}{\bibinfo{person}{Tianqi Chen} {and} \bibinfo{person}{Carlos
  Guestrin}.} \bibinfo{year}{2016}\natexlab{}.
\newblock \showarticletitle{Xgboost: A scalable tree boosting system}. In
  \bibinfo{booktitle}{\emph{Proceedings of KDD}}.
\newblock


\bibitem[\protect\citeauthoryear{Cheng, Bernstein, Danescu-Niculescu-Mizil, and
  Leskovec}{Cheng et~al\mbox{.}}{2017}]%
        {cheng2017anyone}
\bibfield{author}{\bibinfo{person}{Justin Cheng}, \bibinfo{person}{Michael
  Bernstein}, \bibinfo{person}{Cristian Danescu-Niculescu-Mizil}, {and}
  \bibinfo{person}{Jure Leskovec}.} \bibinfo{year}{2017}\natexlab{}.
\newblock \showarticletitle{Anyone can become a troll: Causes of trolling
  behavior in online discussions}. In \bibinfo{booktitle}{\emph{Proceedings of
  CSCW}}. ACM, \bibinfo{pages}{1217--1230}.
\newblock


\bibitem[\protect\citeauthoryear{Choi, Aiello, Varga, and Quercia}{Choi
  et~al\mbox{.}}{2020}]%
        {choi2020ten}
\bibfield{author}{\bibinfo{person}{Minje Choi}, \bibinfo{person}{Luca~Maria
  Aiello}, \bibinfo{person}{Kriszti{\'a}n~Zsolt Varga}, {and}
  \bibinfo{person}{Daniele Quercia}.} \bibinfo{year}{2020}\natexlab{}.
\newblock \showarticletitle{Ten social dimensions of conversations and
  relationships}. In \bibinfo{booktitle}{\emph{Proceedings of The Web
  Conference}}. \bibinfo{pages}{1514--1525}.
\newblock


\bibitem[\protect\citeauthoryear{Danescu-Niculescu-Mizil, Lee, Pang, and
  Kleinberg}{Danescu-Niculescu-Mizil et~al\mbox{.}}{2012}]%
        {danescu2012echoes}
\bibfield{author}{\bibinfo{person}{Cristian Danescu-Niculescu-Mizil},
  \bibinfo{person}{Lillian Lee}, \bibinfo{person}{Bo Pang}, {and}
  \bibinfo{person}{Jon Kleinberg}.} \bibinfo{year}{2012}\natexlab{}.
\newblock \showarticletitle{Echoes of power: Language effects and power
  differences in social interaction}. In \bibinfo{booktitle}{\emph{Proceedings
  of WWW}}. ACM, \bibinfo{pages}{699--708}.
\newblock


\bibitem[\protect\citeauthoryear{Danescu-Niculescu-Mizil, Sudhof, Jurafsky,
  Leskovec, and Potts}{Danescu-Niculescu-Mizil et~al\mbox{.}}{2013}]%
        {danescu2013computational}
\bibfield{author}{\bibinfo{person}{Cristian Danescu-Niculescu-Mizil},
  \bibinfo{person}{Moritz Sudhof}, \bibinfo{person}{Dan Jurafsky},
  \bibinfo{person}{Jure Leskovec}, {and} \bibinfo{person}{Christopher Potts}.}
  \bibinfo{year}{2013}\natexlab{}.
\newblock \showarticletitle{A computational approach to politeness with
  application to social factors}. In \bibinfo{booktitle}{\emph{Proceedings of
  ACL}}.
\newblock


\bibitem[\protect\citeauthoryear{Deri, Rappaz, Aiello, and Quercia}{Deri
  et~al\mbox{.}}{2018}]%
        {deri2018coloring}
\bibfield{author}{\bibinfo{person}{Sebastian Deri}, \bibinfo{person}{Jeremie
  Rappaz}, \bibinfo{person}{Luca~Maria Aiello}, {and} \bibinfo{person}{Daniele
  Quercia}.} \bibinfo{year}{2018}\natexlab{}.
\newblock \showarticletitle{Coloring in the links: Capturing social ties as
  they are perceived}.
\newblock \bibinfo{journal}{\emph{Proceedings of CSCW}} (\bibinfo{year}{2018}).
\newblock


\bibitem[\protect\citeauthoryear{Devlin, Chang, Lee, and Toutanova}{Devlin
  et~al\mbox{.}}{2019}]%
        {devlin2018bert}
\bibfield{author}{\bibinfo{person}{Jacob Devlin}, \bibinfo{person}{Ming-Wei
  Chang}, \bibinfo{person}{Kenton Lee}, {and} \bibinfo{person}{Kristina
  Toutanova}.} \bibinfo{year}{2019}\natexlab{}.
\newblock \showarticletitle{Bert: Pre-training of deep bidirectional
  transformers for language understanding}. In
  \bibinfo{booktitle}{\emph{Proceedings of NAACL}}.
\newblock


\bibitem[\protect\citeauthoryear{Duck}{Duck}{2007}]%
        {duck2007human}
\bibfield{author}{\bibinfo{person}{Steve Duck}.}
  \bibinfo{year}{2007}\natexlab{}.
\newblock \bibinfo{booktitle}{\emph{Human relationships}}.
\newblock \bibinfo{publisher}{Sage}.
\newblock


\bibitem[\protect\citeauthoryear{Duggan}{Duggan}{2017}]%
        {duggan2017online}
\bibfield{author}{\bibinfo{person}{Maeve Duggan}.}
  \bibinfo{year}{2017}\natexlab{}.
\newblock \showarticletitle{Online harassment 2017}.
\newblock  (\bibinfo{year}{2017}).
\newblock


\bibitem[\protect\citeauthoryear{Emmons and McCullough}{Emmons and
  McCullough}{2004}]%
        {emmons2004psychology}
\bibfield{author}{\bibinfo{person}{Robert~A Emmons} {and}
  \bibinfo{person}{Michael~E McCullough}.} \bibinfo{year}{2004}\natexlab{}.
\newblock \bibinfo{booktitle}{\emph{The psychology of gratitude}}.
\newblock \bibinfo{publisher}{Oxford University Press}.
\newblock


\bibitem[\protect\citeauthoryear{{Facebook Research}}{{Facebook
  Research}}{2019}]%
        {instagram2019well}
\bibfield{author}{\bibinfo{person}{{Facebook Research}}.}
  \bibinfo{year}{2019}\natexlab{}.
\newblock \bibinfo{title}{Instagram Request for Proposals for Well-being and
  Safety Research}.
\newblock
\newblock
\newblock
\shownote{\url{https://research.fb.com/programs/research-awards/proposals/instagram-request-for-proposals-for-well-being-and-safety-research/}.}


\bibitem[\protect\citeauthoryear{Fortuna and Nunes}{Fortuna and Nunes}{2018}]%
        {fortuna2018survey}
\bibfield{author}{\bibinfo{person}{Paula Fortuna} {and}
  \bibinfo{person}{S{\'e}rgio Nunes}.} \bibinfo{year}{2018}\natexlab{}.
\newblock \showarticletitle{A survey on automatic detection of hate speech in
  text}.
\newblock \bibinfo{journal}{\emph{ACM Computing Surveys (CSUR)}}
  \bibinfo{volume}{51}, \bibinfo{number}{4} (\bibinfo{year}{2018}),
  \bibinfo{pages}{85}.
\newblock


\bibitem[\protect\citeauthoryear{Frimer, Aquino, Gebauer, Zhu, and
  Oakes}{Frimer et~al\mbox{.}}{2015}]%
        {frimer2015decline}
\bibfield{author}{\bibinfo{person}{Jeremy~A Frimer}, \bibinfo{person}{Karl
  Aquino}, \bibinfo{person}{Jochen~E Gebauer}, \bibinfo{person}{Luke~Lei Zhu},
  {and} \bibinfo{person}{Harrison Oakes}.} \bibinfo{year}{2015}\natexlab{}.
\newblock \showarticletitle{A decline in prosocial language helps explain
  public disapproval of the US Congress}.
\newblock \bibinfo{journal}{\emph{Proceedings of the National Academy of
  Sciences}} \bibinfo{volume}{112}, \bibinfo{number}{21}
  (\bibinfo{year}{2015}), \bibinfo{pages}{6591--6594}.
\newblock


\bibitem[\protect\citeauthoryear{Frimer, Schaefer, and Oakes}{Frimer
  et~al\mbox{.}}{2014}]%
        {frimer2014moral}
\bibfield{author}{\bibinfo{person}{Jeremy~A Frimer}, \bibinfo{person}{Nicola~K
  Schaefer}, {and} \bibinfo{person}{Harrison Oakes}.}
  \bibinfo{year}{2014}\natexlab{}.
\newblock \showarticletitle{Moral actor, selfish agent.}
\newblock \bibinfo{journal}{\emph{Journal of personality and social
  psychology}} \bibinfo{volume}{106}, \bibinfo{number}{5}
  (\bibinfo{year}{2014}), \bibinfo{pages}{790}.
\newblock


\bibitem[\protect\citeauthoryear{Gillespie}{Gillespie}{2018}]%
        {gillespie2018custodians}
\bibfield{author}{\bibinfo{person}{Tarleton Gillespie}.}
  \bibinfo{year}{2018}\natexlab{}.
\newblock \bibinfo{booktitle}{\emph{Custodians of the Internet: Platforms,
  content moderation, and the hidden decisions that shape social media}}.
\newblock \bibinfo{publisher}{Yale University Press}.
\newblock


\bibitem[\protect\citeauthoryear{Goette, Huffman, and Meier}{Goette
  et~al\mbox{.}}{2012}]%
        {goette2012impact}
\bibfield{author}{\bibinfo{person}{Lorenz Goette}, \bibinfo{person}{David
  Huffman}, {and} \bibinfo{person}{Stephan Meier}.}
  \bibinfo{year}{2012}\natexlab{}.
\newblock \showarticletitle{The impact of social ties on group interactions:
  Evidence from minimal groups and randomly assigned real groups}.
\newblock \bibinfo{journal}{\emph{American Economic Journal: Microeconomics}}
  \bibinfo{volume}{4}, \bibinfo{number}{1} (\bibinfo{year}{2012}),
  \bibinfo{pages}{101--15}.
\newblock


\bibitem[\protect\citeauthoryear{Greatbatch and Clark}{Greatbatch and
  Clark}{2003}]%
        {greatbatch2003displaying}
\bibfield{author}{\bibinfo{person}{David Greatbatch} {and}
  \bibinfo{person}{Timothy Clark}.} \bibinfo{year}{2003}\natexlab{}.
\newblock \showarticletitle{Displaying group cohesiveness: Humour and laughter
  in the public lectures of management gurus}.
\newblock \bibinfo{journal}{\emph{Human relations}} \bibinfo{volume}{56},
  \bibinfo{number}{12} (\bibinfo{year}{2003}), \bibinfo{pages}{1515--1544}.
\newblock


\bibitem[\protect\citeauthoryear{Gr{\"o}ndahl, Pajola, Juuti, Conti, and
  Asokan}{Gr{\"o}ndahl et~al\mbox{.}}{2018}]%
        {grondahl2018all}
\bibfield{author}{\bibinfo{person}{Tommi Gr{\"o}ndahl}, \bibinfo{person}{Luca
  Pajola}, \bibinfo{person}{Mika Juuti}, \bibinfo{person}{Mauro Conti}, {and}
  \bibinfo{person}{N Asokan}.} \bibinfo{year}{2018}\natexlab{}.
\newblock \showarticletitle{All You Need is" Love" Evading Hate Speech
  Detection}. In \bibinfo{booktitle}{\emph{Proceedings of the 11th ACM Workshop
  on Artificial Intelligence and Security}}. \bibinfo{pages}{2--12}.
\newblock


\bibitem[\protect\citeauthoryear{Hutto and Gilbert}{Hutto and Gilbert}{2014}]%
        {hutto2014vader}
\bibfield{author}{\bibinfo{person}{Clayton~J Hutto} {and} \bibinfo{person}{Eric
  Gilbert}.} \bibinfo{year}{2014}\natexlab{}.
\newblock \showarticletitle{Vader: A parsimonious rule-based model for
  sentiment analysis of social media text}. In
  \bibinfo{booktitle}{\emph{Proceedings of ICWSM}}.
\newblock


\bibitem[\protect\citeauthoryear{Jiao, Li, Wu, and Mei}{Jiao
  et~al\mbox{.}}{2018}]%
        {jiao2018find}
\bibfield{author}{\bibinfo{person}{Yunhao Jiao}, \bibinfo{person}{Cheng Li},
  \bibinfo{person}{Fei Wu}, {and} \bibinfo{person}{Qiaozhu Mei}.}
  \bibinfo{year}{2018}\natexlab{}.
\newblock \showarticletitle{Find the conversation killers: A predictive study
  of thread-ending posts}. In \bibinfo{booktitle}{\emph{Proceedings of the Web
  Conference}}.
\newblock


\bibitem[\protect\citeauthoryear{Joinson}{Joinson}{2001}]%
        {joinson2001self}
\bibfield{author}{\bibinfo{person}{Adam~N Joinson}.}
  \bibinfo{year}{2001}\natexlab{}.
\newblock \showarticletitle{Self-disclosure in computer-mediated communication:
  The role of self-awareness and visual anonymity}.
\newblock \bibinfo{journal}{\emph{European journal of social psychology}}
  \bibinfo{volume}{31}, \bibinfo{number}{2} (\bibinfo{year}{2001}),
  \bibinfo{pages}{177--192}.
\newblock


\bibitem[\protect\citeauthoryear{Joinson and Paine}{Joinson and Paine}{2007}]%
        {joinson2007self}
\bibfield{author}{\bibinfo{person}{Adam~N Joinson} {and}
  \bibinfo{person}{Carina~B Paine}.} \bibinfo{year}{2007}\natexlab{}.
\newblock \showarticletitle{Self-disclosure, privacy and the Internet}.
\newblock \bibinfo{journal}{\emph{The Oxford handbook of Internet psychology}}
  \bibinfo{volume}{2374252} (\bibinfo{year}{2007}).
\newblock


\bibitem[\protect\citeauthoryear{Joward}{Joward}{1971}]%
        {joward1971self}
\bibfield{author}{\bibinfo{person}{Sidney~M Joward}.}
  \bibinfo{year}{1971}\natexlab{}.
\newblock \bibinfo{booktitle}{\emph{Self-dislosure: An Experimental Analysis of
  the Transparent Self}}.
\newblock \bibinfo{publisher}{Wiley Interscience}.
\newblock


\bibitem[\protect\citeauthoryear{Jurgens, Chandrasekharan, and
  Hemphill}{Jurgens et~al\mbox{.}}{2019}]%
        {jurgens2019just}
\bibfield{author}{\bibinfo{person}{David Jurgens}, \bibinfo{person}{Eshwar
  Chandrasekharan}, {and} \bibinfo{person}{Libby Hemphill}.}
  \bibinfo{year}{2019}\natexlab{}.
\newblock \showarticletitle{A Just and Comprehensive Strategy for Using NLP to
  Address Online Abuse}. In \bibinfo{booktitle}{\emph{Proceedings of the 57th
  Annual Meeting of the Association for Computational Linguistics (ACL)}}.
\newblock


\bibitem[\protect\citeauthoryear{Knickerbocker}{Knickerbocker}{2003}]%
        {knickerbocker2003prosocial}
\bibfield{author}{\bibinfo{person}{Roberta~L Knickerbocker}.}
  \bibinfo{year}{2003}\natexlab{}.
\newblock \showarticletitle{Prosocial behavior}.
\newblock \bibinfo{journal}{\emph{Center on Philanthropy at Indiana
  University}} (\bibinfo{year}{2003}), \bibinfo{pages}{1--3}.
\newblock


\bibitem[\protect\citeauthoryear{Kolhatkar and Taboada}{Kolhatkar and
  Taboada}{2017}]%
        {kolhatkar2017constructive}
\bibfield{author}{\bibinfo{person}{Varada Kolhatkar} {and}
  \bibinfo{person}{Maite Taboada}.} \bibinfo{year}{2017}\natexlab{}.
\newblock \showarticletitle{Constructive language in news comments}. In
  \bibinfo{booktitle}{\emph{Proceedings of the Workshop on Abusive Language
  Online}}.
\newblock


\bibitem[\protect\citeauthoryear{Kolhatkar, Thain, Sorensen, Dixon, and
  Taboada}{Kolhatkar et~al\mbox{.}}{2020}]%
        {kolhatkar2020classifying}
\bibfield{author}{\bibinfo{person}{Varada Kolhatkar}, \bibinfo{person}{Nithum
  Thain}, \bibinfo{person}{Jeffrey Sorensen}, \bibinfo{person}{Lucas Dixon},
  {and} \bibinfo{person}{Maite Taboada}.} \bibinfo{year}{2020}\natexlab{}.
\newblock \showarticletitle{Classifying Constructive Comments}.
\newblock \bibinfo{journal}{\emph{First Monday.}} (\bibinfo{year}{2020}).
\newblock


\bibitem[\protect\citeauthoryear{Kulesza, Dolinski, Huisman, and
  Majewski}{Kulesza et~al\mbox{.}}{2014}]%
        {kulesza2014echo}
\bibfield{author}{\bibinfo{person}{Wojciech Kulesza}, \bibinfo{person}{Dariusz
  Dolinski}, \bibinfo{person}{Avia Huisman}, {and} \bibinfo{person}{Robert
  Majewski}.} \bibinfo{year}{2014}\natexlab{}.
\newblock \showarticletitle{The echo effect: The power of verbal mimicry to
  influence prosocial behavior}.
\newblock \bibinfo{journal}{\emph{Journal of Language and Social Psychology}}
  \bibinfo{volume}{33}, \bibinfo{number}{2} (\bibinfo{year}{2014}),
  \bibinfo{pages}{183--201}.
\newblock


\bibitem[\protect\citeauthoryear{Kumar, Cheng, and Leskovec}{Kumar
  et~al\mbox{.}}{2017}]%
        {kumar2017antisocial}
\bibfield{author}{\bibinfo{person}{Srijan Kumar}, \bibinfo{person}{Justin
  Cheng}, {and} \bibinfo{person}{Jure Leskovec}.}
  \bibinfo{year}{2017}\natexlab{}.
\newblock \showarticletitle{Antisocial behavior on the web: Characterization
  and detection}. In \bibinfo{booktitle}{\emph{Proceedings of the 26th
  International Conference on World Wide Web Companion}}.
  \bibinfo{pages}{947--950}.
\newblock


\bibitem[\protect\citeauthoryear{Kumar, Zhang, and Leskovec}{Kumar
  et~al\mbox{.}}{2019}]%
        {kumar2019predicting}
\bibfield{author}{\bibinfo{person}{Srijan Kumar}, \bibinfo{person}{Xikun
  Zhang}, {and} \bibinfo{person}{Jure Leskovec}.}
  \bibinfo{year}{2019}\natexlab{}.
\newblock \showarticletitle{Predicting dynamic embedding trajectory in temporal
  interaction networks}. In \bibinfo{booktitle}{\emph{Proceedings of KD}}. ACM,
  \bibinfo{pages}{1269--1278}.
\newblock


\bibitem[\protect\citeauthoryear{Lan, Chen, Goodman, Gimpel, Sharma, and
  Soricut}{Lan et~al\mbox{.}}{2019}]%
        {lan2019albert}
\bibfield{author}{\bibinfo{person}{Zhenzhong Lan}, \bibinfo{person}{Mingda
  Chen}, \bibinfo{person}{Sebastian Goodman}, \bibinfo{person}{Kevin Gimpel},
  \bibinfo{person}{Piyush Sharma}, {and} \bibinfo{person}{Radu Soricut}.}
  \bibinfo{year}{2019}\natexlab{}.
\newblock \showarticletitle{Albert: A lite bert for self-supervised learning of
  language representations}.
\newblock \bibinfo{journal}{\emph{arXiv preprint arXiv:1909.11942}}
  (\bibinfo{year}{2019}).
\newblock


\bibitem[\protect\citeauthoryear{Lazer}{Lazer}{2015}]%
        {lazer2015rise}
\bibfield{author}{\bibinfo{person}{David Lazer}.}
  \bibinfo{year}{2015}\natexlab{}.
\newblock \showarticletitle{The rise of the social algorithm}.
\newblock \bibinfo{journal}{\emph{Science}} \bibinfo{volume}{348},
  \bibinfo{number}{6239} (\bibinfo{year}{2015}), \bibinfo{pages}{1090--1091}.
\newblock


\bibitem[\protect\citeauthoryear{Liu, Guberman, Hemphill, and Culotta}{Liu
  et~al\mbox{.}}{2018}]%
        {liu2018forecasting}
\bibfield{author}{\bibinfo{person}{Ping Liu}, \bibinfo{person}{Joshua
  Guberman}, \bibinfo{person}{Libby Hemphill}, {and} \bibinfo{person}{Aron
  Culotta}.} \bibinfo{year}{2018}\natexlab{}.
\newblock \showarticletitle{Forecasting the presence and intensity of hostility
  on Instagram using linguistic and social features}. In
  \bibinfo{booktitle}{\emph{Proceedings of ICWSM}}.
\newblock


\bibitem[\protect\citeauthoryear{Matias}{Matias}{2019}]%
        {matias2019preventing}
\bibfield{author}{\bibinfo{person}{J~Nathan Matias}.}
  \bibinfo{year}{2019}\natexlab{}.
\newblock \showarticletitle{Preventing harassment and increasing group
  participation through social norms in 2,190 online science discussions}.
\newblock \bibinfo{journal}{\emph{Proceedings of the National Academy of
  Sciences}} \bibinfo{volume}{116}, \bibinfo{number}{20}
  (\bibinfo{year}{2019}), \bibinfo{pages}{9785--9789}.
\newblock


\bibitem[\protect\citeauthoryear{McCallum}{McCallum}{2002}]%
        {McCallumMALLET}
\bibfield{author}{\bibinfo{person}{Andrew~Kachites McCallum}.}
  \bibinfo{year}{2002}\natexlab{}.
\newblock \bibinfo{title}{MALLET: A Machine Learning for Language Toolkit}.
  (\bibinfo{year}{2002}).
\newblock
\newblock
\shownote{http://mallet.cs.umass.edu.}


\bibitem[\protect\citeauthoryear{McCullough, Kilpatrick, Emmons, and
  Larson}{McCullough et~al\mbox{.}}{2001}]%
        {mccullough2001gratitude}
\bibfield{author}{\bibinfo{person}{Michael~E McCullough},
  \bibinfo{person}{Shelley~D Kilpatrick}, \bibinfo{person}{Robert~A Emmons},
  {and} \bibinfo{person}{David~B Larson}.} \bibinfo{year}{2001}\natexlab{}.
\newblock \showarticletitle{Is gratitude a moral affect?}
\newblock \bibinfo{journal}{\emph{Psychological bulletin}}
  \bibinfo{volume}{127}, \bibinfo{number}{2} (\bibinfo{year}{2001}),
  \bibinfo{pages}{249}.
\newblock


\bibitem[\protect\citeauthoryear{McCullough, Kimeldorf, and Cohen}{McCullough
  et~al\mbox{.}}{2008}]%
        {mccullough2008adaptation}
\bibfield{author}{\bibinfo{person}{Michael~E McCullough},
  \bibinfo{person}{Marcia~B Kimeldorf}, {and} \bibinfo{person}{Adam~D Cohen}.}
  \bibinfo{year}{2008}\natexlab{}.
\newblock \showarticletitle{An adaptation for altruism: The social causes,
  social effects, and social evolution of gratitude}.
\newblock \bibinfo{journal}{\emph{Current directions in psychological science}}
  \bibinfo{volume}{17}, \bibinfo{number}{4} (\bibinfo{year}{2008}),
  \bibinfo{pages}{281--285}.
\newblock


\bibitem[\protect\citeauthoryear{Milne, Pink, Hachey, and Calvo}{Milne
  et~al\mbox{.}}{2016}]%
        {milne2016clpsych}
\bibfield{author}{\bibinfo{person}{David~N Milne}, \bibinfo{person}{Glen Pink},
  \bibinfo{person}{Ben Hachey}, {and} \bibinfo{person}{Rafael~A Calvo}.}
  \bibinfo{year}{2016}\natexlab{}.
\newblock \showarticletitle{Clpsych 2016 shared task: Triaging content in
  online peer-support forums}. In \bibinfo{booktitle}{\emph{Proceedings of the
  Third Workshop on Computational Linguistics and Clinical Psychology}}.
  \bibinfo{pages}{118--127}.
\newblock


\bibitem[\protect\citeauthoryear{Mussen and Eisenberg-Berg}{Mussen and
  Eisenberg-Berg}{1977}]%
        {mussen1977roots}
\bibfield{author}{\bibinfo{person}{Paul Mussen} {and} \bibinfo{person}{Nancy
  Eisenberg-Berg}.} \bibinfo{year}{1977}\natexlab{}.
\newblock \bibinfo{booktitle}{\emph{Roots of caring, sharing, and helping: The
  development of pro-social behavior in children.}}
\newblock \bibinfo{publisher}{WH Freeman}.
\newblock


\bibitem[\protect\citeauthoryear{Napoles, Pappu, and Tetreault}{Napoles
  et~al\mbox{.}}{2017a}]%
        {napoles2017automatically}
\bibfield{author}{\bibinfo{person}{Courtney Napoles}, \bibinfo{person}{Aasish
  Pappu}, {and} \bibinfo{person}{Joel Tetreault}.}
  \bibinfo{year}{2017}\natexlab{a}.
\newblock \showarticletitle{Automatically identifying good conversations online
  (yes, they do exist!)}. In \bibinfo{booktitle}{\emph{Proceedings of ICWSM}}.
\newblock


\bibitem[\protect\citeauthoryear{Napoles, Tetreault, Pappu, Rosato, and
  Provenzale}{Napoles et~al\mbox{.}}{2017b}]%
        {napoles2017finding}
\bibfield{author}{\bibinfo{person}{Courtney Napoles}, \bibinfo{person}{Joel
  Tetreault}, \bibinfo{person}{Aasish Pappu}, \bibinfo{person}{Enrica Rosato},
  {and} \bibinfo{person}{Brian Provenzale}.} \bibinfo{year}{2017}\natexlab{b}.
\newblock \showarticletitle{Finding good conversations online: The Yahoo News
  annotated comments corpus}. In \bibinfo{booktitle}{\emph{Proceedings of the
  11th Linguistic Annotation Workshop}}.
\newblock


\bibitem[\protect\citeauthoryear{Navindgi, Brun, Masson, and Nowson}{Navindgi
  et~al\mbox{.}}{2016}]%
        {navindgi2016steps}
\bibfield{author}{\bibinfo{person}{Amit Navindgi}, \bibinfo{person}{Caroline
  Brun}, \bibinfo{person}{C{\'e}cile~Boulard Masson}, {and}
  \bibinfo{person}{Scott Nowson}.} \bibinfo{year}{2016}\natexlab{}.
\newblock \showarticletitle{Steps toward automatic understanding of the
  function of affective language in support groups}. In
  \bibinfo{booktitle}{\emph{Proceedings of The Fourth International Workshop on
  Natural Language Processing for Social Media}}. \bibinfo{pages}{26--33}.
\newblock


\bibitem[\protect\citeauthoryear{Niederhoffer and Pennebaker}{Niederhoffer and
  Pennebaker}{2002}]%
        {niederhoffer2002linguistic}
\bibfield{author}{\bibinfo{person}{Kate~G Niederhoffer} {and}
  \bibinfo{person}{James~W Pennebaker}.} \bibinfo{year}{2002}\natexlab{}.
\newblock \showarticletitle{Linguistic style matching in social interaction}.
\newblock \bibinfo{journal}{\emph{Journal of Language and Social Psychology}}
  \bibinfo{volume}{21}, \bibinfo{number}{4} (\bibinfo{year}{2002}),
  \bibinfo{pages}{337--360}.
\newblock


\bibitem[\protect\citeauthoryear{Orvell, Kross, and Gelman}{Orvell
  et~al\mbox{.}}{2017}]%
        {orvell2017you}
\bibfield{author}{\bibinfo{person}{Ariana Orvell}, \bibinfo{person}{Ethan
  Kross}, {and} \bibinfo{person}{Susan~A Gelman}.}
  \bibinfo{year}{2017}\natexlab{}.
\newblock \showarticletitle{How “you” makes meaning}.
\newblock \bibinfo{journal}{\emph{Science}} \bibinfo{volume}{355},
  \bibinfo{number}{6331} (\bibinfo{year}{2017}), \bibinfo{pages}{1299--1302}.
\newblock


\bibitem[\protect\citeauthoryear{Owren and Bachorowski}{Owren and
  Bachorowski}{2003}]%
        {owren2003reconsidering}
\bibfield{author}{\bibinfo{person}{Michael~J Owren} {and}
  \bibinfo{person}{Jo-Anne Bachorowski}.} \bibinfo{year}{2003}\natexlab{}.
\newblock \showarticletitle{Reconsidering the evolution of nonlinguistic
  communication: The case of laughter}.
\newblock \bibinfo{journal}{\emph{Journal of Nonverbal Behavior}}
  \bibinfo{volume}{27}, \bibinfo{number}{3} (\bibinfo{year}{2003}),
  \bibinfo{pages}{183--200}.
\newblock


\bibitem[\protect\citeauthoryear{Prinstein and Cillessen}{Prinstein and
  Cillessen}{2003}]%
        {prinstein2003forms}
\bibfield{author}{\bibinfo{person}{Mitchell~J Prinstein} {and}
  \bibinfo{person}{Antonius~HN Cillessen}.} \bibinfo{year}{2003}\natexlab{}.
\newblock \showarticletitle{Forms and functions of adolescent peer aggression
  associated with high levels of peer status}.
\newblock \bibinfo{journal}{\emph{Merrill-Palmer Quarterly (1982-)}}
  (\bibinfo{year}{2003}), \bibinfo{pages}{310--342}.
\newblock


\bibitem[\protect\citeauthoryear{Roberts}{Roberts}{2014}]%
        {roberts2014behind}
\bibfield{author}{\bibinfo{person}{Sarah~T Roberts}.}
  \bibinfo{year}{2014}\natexlab{}.
\newblock \emph{\bibinfo{title}{Behind the screen: The hidden digital labor of
  commercial content moderation}}.
\newblock \bibinfo{thesistype}{Ph.D. Dissertation}. \bibinfo{school}{University
  of Illinois at Urbana-Champaign}.
\newblock


\bibitem[\protect\citeauthoryear{Saha, Chandrasekharan, and De~Choudhury}{Saha
  et~al\mbox{.}}{2019}]%
        {saha2019prevalence}
\bibfield{author}{\bibinfo{person}{Koustuv Saha}, \bibinfo{person}{Eshwar
  Chandrasekharan}, {and} \bibinfo{person}{Munmun De~Choudhury}.}
  \bibinfo{year}{2019}\natexlab{}.
\newblock \showarticletitle{Prevalence and Psychological Effects of Hateful
  Speech in Online College Communities.}. In
  \bibinfo{booktitle}{\emph{Proceedings of Web Science}}.
\newblock


\bibitem[\protect\citeauthoryear{Sap, Card, Gabriel, Choi, and Smith}{Sap
  et~al\mbox{.}}{2019}]%
        {sap2019risk}
\bibfield{author}{\bibinfo{person}{Maarten Sap}, \bibinfo{person}{Dallas Card},
  \bibinfo{person}{Saadia Gabriel}, \bibinfo{person}{Yejin Choi}, {and}
  \bibinfo{person}{Noah~A Smith}.} \bibinfo{year}{2019}\natexlab{}.
\newblock \showarticletitle{The risk of racial bias in hate speech detection}.
  In \bibinfo{booktitle}{\emph{Proceedings of ACL}}.
\newblock


\bibitem[\protect\citeauthoryear{Scissors, Gill, and Gergle}{Scissors
  et~al\mbox{.}}{2008}]%
        {scissors2008linguistic}
\bibfield{author}{\bibinfo{person}{Lauren~E Scissors},
  \bibinfo{person}{Alastair~J Gill}, {and} \bibinfo{person}{Darren Gergle}.}
  \bibinfo{year}{2008}\natexlab{}.
\newblock \showarticletitle{Linguistic mimicry and trust in text-based {CMC}}.
  In \bibinfo{booktitle}{\emph{Proceedings of CSCW}}.
\newblock


\bibitem[\protect\citeauthoryear{Sharma, Miner, Atkins, and Althoff}{Sharma
  et~al\mbox{.}}{2020}]%
        {sharma2020computational}
\bibfield{author}{\bibinfo{person}{Ashish Sharma}, \bibinfo{person}{Adam~S
  Miner}, \bibinfo{person}{David~C Atkins}, {and} \bibinfo{person}{Tim
  Althoff}.} \bibinfo{year}{2020}\natexlab{}.
\newblock \showarticletitle{A Computational Approach to Understanding Empathy
  Expressed in Text-Based Mental Health Support}. In
  \bibinfo{booktitle}{\emph{Proceedings of EMNLP}}.
\newblock


\bibitem[\protect\citeauthoryear{Sproull}{Sproull}{2011}]%
        {sproull2011prosocial}
\bibfield{author}{\bibinfo{person}{Lee Sproull}.}
  \bibinfo{year}{2011}\natexlab{}.
\newblock \showarticletitle{Prosocial behavior on the net}.
\newblock \bibinfo{journal}{\emph{Daedalus}} \bibinfo{volume}{140},
  \bibinfo{number}{4} (\bibinfo{year}{2011}), \bibinfo{pages}{140--153}.
\newblock


\bibitem[\protect\citeauthoryear{Taylor and Thomas}{Taylor and Thomas}{2008}]%
        {taylor2008linguistic}
\bibfield{author}{\bibinfo{person}{Paul~J Taylor} {and} \bibinfo{person}{Sally
  Thomas}.} \bibinfo{year}{2008}\natexlab{}.
\newblock \showarticletitle{Linguistic style matching and negotiation outcome}.
\newblock \bibinfo{journal}{\emph{Negotiation and Conflict Management
  Research}} \bibinfo{volume}{1}, \bibinfo{number}{3} (\bibinfo{year}{2008}),
  \bibinfo{pages}{263--281}.
\newblock


\bibitem[\protect\citeauthoryear{Twenge, Baumeister, DeWall, Ciarocco, and
  Bartels}{Twenge et~al\mbox{.}}{2007}]%
        {twenge2007social}
\bibfield{author}{\bibinfo{person}{Jean~M Twenge}, \bibinfo{person}{Roy~F
  Baumeister}, \bibinfo{person}{C~Nathan DeWall}, \bibinfo{person}{Natalie~J
  Ciarocco}, {and} \bibinfo{person}{J~Michael Bartels}.}
  \bibinfo{year}{2007}\natexlab{}.
\newblock \showarticletitle{Social exclusion decreases prosocial behavior.}
\newblock \bibinfo{journal}{\emph{Journal of personality and social
  psychology}} \bibinfo{volume}{92}, \bibinfo{number}{1}
  (\bibinfo{year}{2007}), \bibinfo{pages}{56}.
\newblock


\bibitem[\protect\citeauthoryear{Twitter}{Twitter}{2018}]%
        {twitter2018health}
\bibfield{author}{\bibinfo{person}{Twitter}.} \bibinfo{year}{2018}\natexlab{}.
\newblock \bibinfo{title}{Twitter health metrics proposal submission}.
\newblock
\newblock
\newblock
\shownote{\url{https://blog.twitter.com/en_us/topics/company/2018/twitter-health-metrics-proposal-submission.html}.}


\bibitem[\protect\citeauthoryear{Vidgen, Harris, Nguyen, Tromble, Hale, and
  Margetts}{Vidgen et~al\mbox{.}}{2019}]%
        {vidgen2019challenges}
\bibfield{author}{\bibinfo{person}{Bertie Vidgen}, \bibinfo{person}{Alex
  Harris}, \bibinfo{person}{Dong Nguyen}, \bibinfo{person}{Rebekah Tromble},
  \bibinfo{person}{Scott Hale}, {and} \bibinfo{person}{Helen Margetts}.}
  \bibinfo{year}{2019}\natexlab{}.
\newblock \showarticletitle{Challenges and frontiers in abusive content
  detection}. In \bibinfo{booktitle}{\emph{ACL}}.
\newblock


\bibitem[\protect\citeauthoryear{Voigt, Camp, Prabhakaran, Hamilton, Hetey,
  Griffiths, Jurgens, Jurafsky, and Eberhardt}{Voigt et~al\mbox{.}}{2017}]%
        {voigt2017language}
\bibfield{author}{\bibinfo{person}{Rob Voigt}, \bibinfo{person}{Nicholas~P
  Camp}, \bibinfo{person}{Vinodkumar Prabhakaran}, \bibinfo{person}{William~L
  Hamilton}, \bibinfo{person}{Rebecca~C Hetey}, \bibinfo{person}{Camilla~M
  Griffiths}, \bibinfo{person}{David Jurgens}, \bibinfo{person}{Dan Jurafsky},
  {and} \bibinfo{person}{Jennifer~L Eberhardt}.}
  \bibinfo{year}{2017}\natexlab{}.
\newblock \showarticletitle{Language from police body camera footage shows
  racial disparities in officer respect}.
\newblock \bibinfo{journal}{\emph{Proceedings of the National Academy of
  Sciences}} \bibinfo{volume}{114}, \bibinfo{number}{25}
  (\bibinfo{year}{2017}), \bibinfo{pages}{6521--6526}.
\newblock


\bibitem[\protect\citeauthoryear{Wang, Yen, and Reitter}{Wang
  et~al\mbox{.}}{2015}]%
        {wang2015pragmatic}
\bibfield{author}{\bibinfo{person}{Yafei Wang}, \bibinfo{person}{John Yen},
  {and} \bibinfo{person}{David Reitter}.} \bibinfo{year}{2015}\natexlab{}.
\newblock \showarticletitle{Pragmatic alignment on social support type in
  health forum conversations}. In \bibinfo{booktitle}{\emph{Proceedings of the
  6th Workshop on Cognitive Modeling and Computational Linguistics}}.
  \bibinfo{pages}{9--18}.
\newblock


\bibitem[\protect\citeauthoryear{Wang and Jurgens}{Wang and Jurgens}{2018}]%
        {wang2018its}
\bibfield{author}{\bibinfo{person}{Zijian Wang} {and} \bibinfo{person}{David
  Jurgens}.} \bibinfo{year}{2018}\natexlab{}.
\newblock \showarticletitle{It's going to be okay: Measuring Access to Support
  in Online Communities}. In \bibinfo{booktitle}{\emph{Proceedings of EMNLP}}.
\newblock


\bibitem[\protect\citeauthoryear{Waseem, Davidson, Warmsley, and Weber}{Waseem
  et~al\mbox{.}}{2017}]%
        {waseem2017understanding}
\bibfield{author}{\bibinfo{person}{Zeerak Waseem}, \bibinfo{person}{Thomas
  Davidson}, \bibinfo{person}{Dana Warmsley}, {and} \bibinfo{person}{Ingmar
  Weber}.} \bibinfo{year}{2017}\natexlab{}.
\newblock \showarticletitle{Understanding abuse: A typology of abusive language
  detection subtasks}. In \bibinfo{booktitle}{\emph{Proceedings of the First
  Workshop on Abusive Language}}.
\newblock


\bibitem[\protect\citeauthoryear{Wills}{Wills}{1991}]%
        {wills1991social}
\bibfield{author}{\bibinfo{person}{Thomas~Ashby Wills}.}
  \bibinfo{year}{1991}\natexlab{}.
\newblock \showarticletitle{Social support and interpersonal relationships.}
\newblock  (\bibinfo{year}{1991}).
\newblock


\bibitem[\protect\citeauthoryear{Wolf, Debut, Sanh, Chaumond, Delangue, Moi,
  Cistac, Rault, Louf, Funtowicz, Davison, Shleifer, von Platen, Ma, Jernite,
  Plu, Xu, Scao, Gugger, Drame, Lhoest, and Rush}{Wolf et~al\mbox{.}}{2020}]%
        {wolf-etal-2020-transformers}
\bibfield{author}{\bibinfo{person}{Thomas Wolf}, \bibinfo{person}{Lysandre
  Debut}, \bibinfo{person}{Victor Sanh}, \bibinfo{person}{Julien Chaumond},
  \bibinfo{person}{Clement Delangue}, \bibinfo{person}{Anthony Moi},
  \bibinfo{person}{Pierric Cistac}, \bibinfo{person}{Tim Rault},
  \bibinfo{person}{Rémi Louf}, \bibinfo{person}{Morgan Funtowicz},
  \bibinfo{person}{Joe Davison}, \bibinfo{person}{Sam Shleifer},
  \bibinfo{person}{Patrick von Platen}, \bibinfo{person}{Clara Ma},
  \bibinfo{person}{Yacine Jernite}, \bibinfo{person}{Julien Plu},
  \bibinfo{person}{Canwen Xu}, \bibinfo{person}{Teven~Le Scao},
  \bibinfo{person}{Sylvain Gugger}, \bibinfo{person}{Mariama Drame},
  \bibinfo{person}{Quentin Lhoest}, {and} \bibinfo{person}{Alexander~M. Rush}.}
  \bibinfo{year}{2020}\natexlab{}.
\newblock \showarticletitle{Transformers: State-of-the-Art Natural Language
  Processing}. In \bibinfo{booktitle}{\emph{Proceedings of EMNLP}}.
\newblock


\bibitem[\protect\citeauthoryear{Wright and Li}{Wright and Li}{2011}]%
        {wright2011associations}
\bibfield{author}{\bibinfo{person}{Michelle~F Wright} {and}
  \bibinfo{person}{Yan Li}.} \bibinfo{year}{2011}\natexlab{}.
\newblock \showarticletitle{The associations between young adults’
  face-to-face prosocial behaviors and their online prosocial behaviors}.
\newblock \bibinfo{journal}{\emph{Computers in Human Behavior}}
  \bibinfo{volume}{27}, \bibinfo{number}{5} (\bibinfo{year}{2011}),
  \bibinfo{pages}{1959--1962}.
\newblock


\bibitem[\protect\citeauthoryear{Wright and Li}{Wright and Li}{2012}]%
        {wright2012prosocial}
\bibfield{author}{\bibinfo{person}{Michelle~F Wright} {and}
  \bibinfo{person}{Yan Li}.} \bibinfo{year}{2012}\natexlab{}.
\newblock \showarticletitle{Prosocial behaviors in the cyber context}.
\newblock In \bibinfo{booktitle}{\emph{Encyclopedia of cyber behavior}}.
  \bibinfo{publisher}{IGI Global}, \bibinfo{pages}{328--341}.
\newblock


\bibitem[\protect\citeauthoryear{Wulczyn, Thain, and Dixon}{Wulczyn
  et~al\mbox{.}}{2017}]%
        {wulczyn2017ex}
\bibfield{author}{\bibinfo{person}{Ellery Wulczyn}, \bibinfo{person}{Nithum
  Thain}, {and} \bibinfo{person}{Lucas Dixon}.}
  \bibinfo{year}{2017}\natexlab{}.
\newblock \showarticletitle{Ex machina: Personal attacks seen at scale}. In
  \bibinfo{booktitle}{\emph{Proceedings of the Web Conference}}.
\newblock


\bibitem[\protect\citeauthoryear{Zhang, Chang, Danescu-Niculescu-Mizil, Dixon,
  Hua, Thain, and Taraborelli}{Zhang et~al\mbox{.}}{2018}]%
        {zhang2018conversations}
\bibfield{author}{\bibinfo{person}{Justine Zhang}, \bibinfo{person}{Jonathan~P
  Chang}, \bibinfo{person}{Cristian Danescu-Niculescu-Mizil},
  \bibinfo{person}{Lucas Dixon}, \bibinfo{person}{Yiqing Hua},
  \bibinfo{person}{Nithum Thain}, {and} \bibinfo{person}{Dario Taraborelli}.}
  \bibinfo{year}{2018}\natexlab{}.
\newblock \showarticletitle{Conversations gone awry: Detecting early signs of
  conversational failure}. In \bibinfo{booktitle}{\emph{Proceedings of ACL}}.
\newblock


\bibitem[\protect\citeauthoryear{Zhou and Jurgens}{Zhou and Jurgens}{2020}]%
        {zhou2020condolences}
\bibfield{author}{\bibinfo{person}{Naitian Zhou} {and} \bibinfo{person}{David
  Jurgens}.} \bibinfo{year}{2020}\natexlab{}.
\newblock \showarticletitle{Condolences and Empathy in Online Communities}. In
  \bibinfo{booktitle}{\emph{Proceedings of EMNLP}}.
\newblock


\end{thebibliography}

\appendix

\section{Prosocial Metrics}

This section describes the features, training, and setup for classifiers and regressors that estimate specific prosocial metrics.

\subsection{Information Sharing }
\label{app:info-share-sec}

Information-sharing comments were identified using a classifier trained on heuristically-labeled data.  Positive examples of  information sharing were drawn from 18 question-focused subreddits where individuals post questions and receive potentially-informative replies (e.g.,  r/whatisthisthing and r/AskReddit); these subreddits cover multiple topics to prevent overfitting to sharing just one type of information. Information sharing comments were drawn from January--March of 2018 from posts that contained at least one question; replies  to these questions  receiving a  score  $>$2 were  taken  as  positive  examples of  information  sharing. Negative  examples were drawn from a random sample of English-language replies to posts \textit{not} in  these  subreddits. Our dataset consists of 55,542 informative comments and 300,226 comments from non-informative communities. This class skew was intentionally left imbalanced to simulate the real-life scenario where most comments are not \textit{information-sharing}.
We fit a logistic regression classifier on the unigram and bigram features with five-fold cross-validation. The hyperparameter for the minimum n-gram frequency was varied between values $[100, 50, 25, 15, 10]$. The final model obtained an F1-score of 0.713. 
We selected a decision threshold of $\ge$0.7 for being information sharing to reduce false positives. 

We further labeled  a comment as information-sharing if it contained a URLs to websites that are commonly used as information references: 
\url{wikipedia.org},
\url{stackoverflow.com},
\url{quora.com},
\url{imdb.com},
\url{webmd.com},
\url{merriam-webster.com},
\url{nih.gov},
\url{weather.com},
\url{genius.com},
\url{books.google.com},
\url{github.com},
\url{wikihow.com},
\url{answers.yahoo.com},
\url{ehow.com},
\url{thefreedictionary.com},
\url{dictionary.com},  and
\url{lifehacker.com}.

\subsection{Laughter}
\label{appendix-laughter}
Laughter is detected by identifying colloquial internet expressions signalling laughter, e.g., ``haha'' or ``lol.'' As these forms may repeat or have variation, e.g., ``hahhahaaha'' we use a regex to detect them:
{
\small \begin{verbatim}
\ba*h+a+h+a+(h+a+)*?h*\b|\bl+o+l+(o+l+)*?\b|\bh+e+h+e+(h+e+)*?h*\b
\end{verbatim} 
}

\subsection{Mentoring}
\label{appendix-mentoring}
We built a classifier with positive examples of mentoring drawn from advice-based subreddits where users post questions and the community responds with advice to those questions (appearing as \tlc). Negative examples were randomly drawn from a \tlc made in all other subreddits.  We considered communities containing the word ``Advice'' in their name (e.g., r/legaladvice, r/relationship\_advice, and r/mechanicadvice), excluding those with the word ``Bad.''. 
Compared to \textit{information-sharing} subreddits, answers in \textit{mentoring} subreddits are typically subjective in nature. We generated 500,000 negative examples using reservoir sampling. We processed and purified this dataset with the same pipeline as \ref{app:info-share-sec}, which resulted in a 79,430 positive examples of mentoring and 299,006 negative examples. 
Our logistic regression classifier for \textit{mentoring} prediction has an F1-score of 0.762 and we manually adjust the decision threshold to $\ge$0.7 in order to reduce false positives.

\subsection{Gratitude}
\label{appendix-gratitude}
To detect gratitude in replies, we use a fixed lexicon of words and phrases, which manual inspection showed had high precision when interpreting whether the responding user was expressing gratitude. Gratitude words are ``thanks,'' ``contented,'' and ``blessed.''  Gratitude phrases are ``thank you,'' ``thankful for,'' ``grateful for,'' ``greatful for,'' ``my gratitude,'' ``i appreciate,'' ``make me smile,'' ``I super appreciate,''  ``i deeply appreciate,'' ``i really appreciate,'' and ``bless your soul.''

\subsection{Esteem Enhancement}
\label{appendix-esteem}
Compliments were identified using a rule-based procedure to select parts of comments referring to the user being replied to and then testing whether the sentiment around that reference was positive. An initial set of candidates was identified by looking for direct mentions of ``you is/are'' or  ``your [word] is/are.'' We then filter out all candidates containing where ``you'' is immediately preceded by ``if'' or ``when'' as analysis showed these constructions were likely to invoke the generic sense of you \cite{berry2009you,orvell2017you} and not refer directly to the user in the parent comment.
From the remaining, we extract the five words following our matched phrase and score the sentiment using VADER \cite{hutto2014vader}. 
We use the compound sentiment score from the \texttt{vaderSentiment} library as an aggregate estimate of the positivity toward the parent comment's user. The minimum threshold for sentiment was set at 0.7 after reviewing several hundred  sentiment scores showed this resulted in few false positives.

\subsection{Donations}
\label{appendix-donating}
Fundraising and Donation behavior is measured  by counting how many times a URL with one of these base domains are mentioned in the total conversation. The following URLs were drawn popular charity, fundraising, and donation organizations:
   \url{gofundme.com},  
\url{indiegogo.com},  
\url{causes.com},  
\url{kickstarter.com},  
\url{patreon.com},  
\url{circleup.com},  
\url{lendingclub.com},  
\url{fundly.com},  
\url{donatekindly.org},  
\url{givecampus.com},  
\url{snap-raise.com},  
\url{snowballfundraising.com},  
\url{bonfire.com},  
\url{crowdrise.com},  
\url{dojiggy.com},  
\url{mightycause.com},  
\url{depositagift.com},  
\url{wemakeit.com},  
\url{donorschoose.org},  
\url{fundrazr.com},  
\url{rallyme.com},  
\url{startsomegood.com},  
\url{diabetes.org},  
\url{humanesociety.org},  
\url{cancer.org},  
\url{nwf.org}, \\
\url{worldwildlife.org},  
\url{habitat.org},  
\url{oxfam.org},  
\url{unicefusa.org},  
\url{wish.org},  
\url{nature.org},  
\url{aspca.org},  
\url{savethechildren.org},  
\url{wfp.org},  
\url{hrc.org},  
\url{hrw.org},  
\url{nationalmssociety.org},  
\url{redcross.org},  
\url{mentalhealthamerica.net},  
\url{amnesty.org},  
\url{heart.org},  
\url{crs.org},  
\url{kiva.org},  
\url{fsf.org},  
\url{rotary.org},  
\url{alz.org}, 
\url{doctorswithoutborders.org},  
\url{unitedway.org},  
 and
\url{cancer.org}.

\begin{figure}[t]
    \centering
    \includegraphics[width=0.46\textwidth]{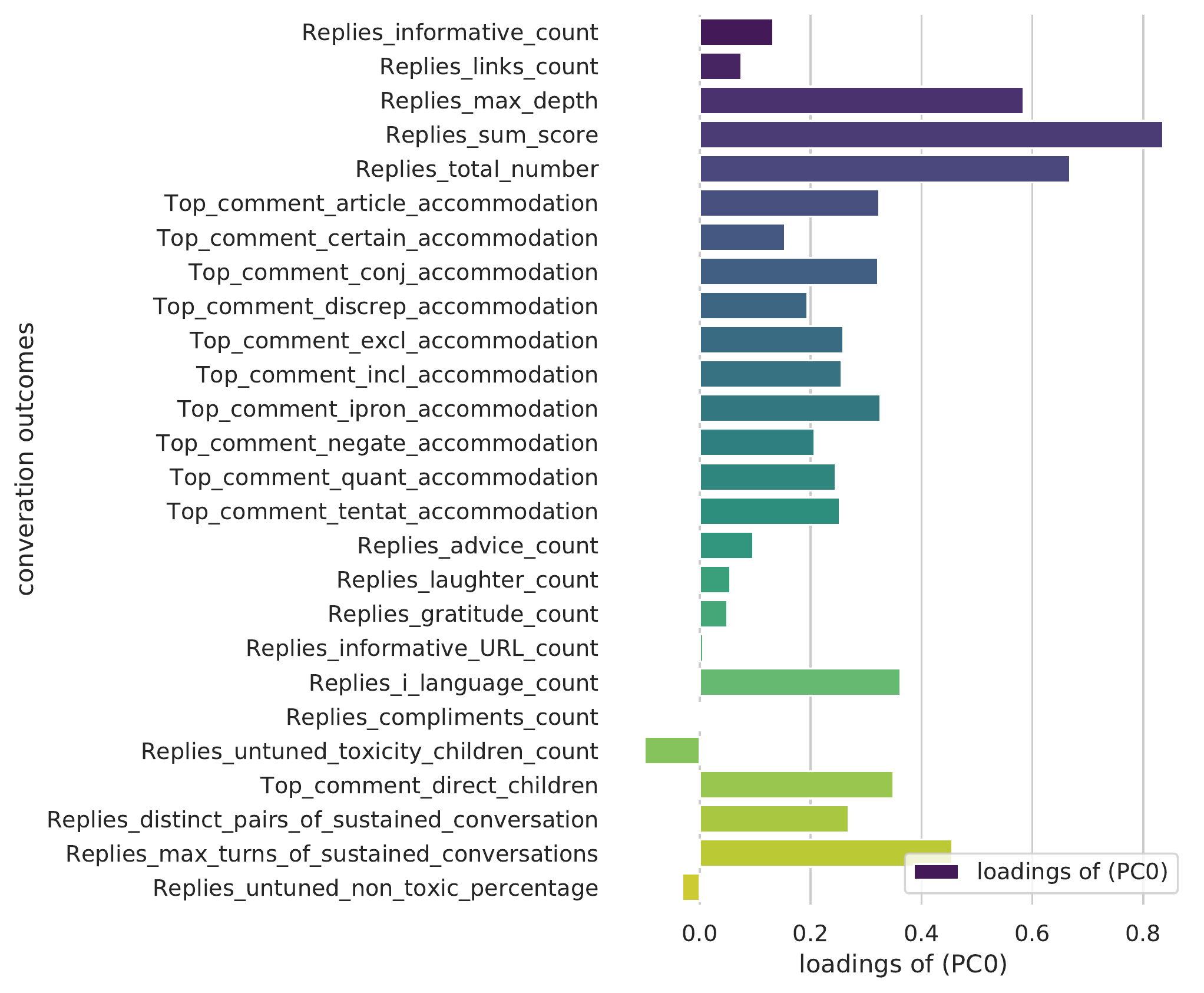}
    \caption{PCA Component 0 Loadings Across Metrics.}
    \label{fig:loadingforpc0}
\end{figure}

\begin{figure}[t]
    \centering
    \includegraphics[width=0.46\textwidth]{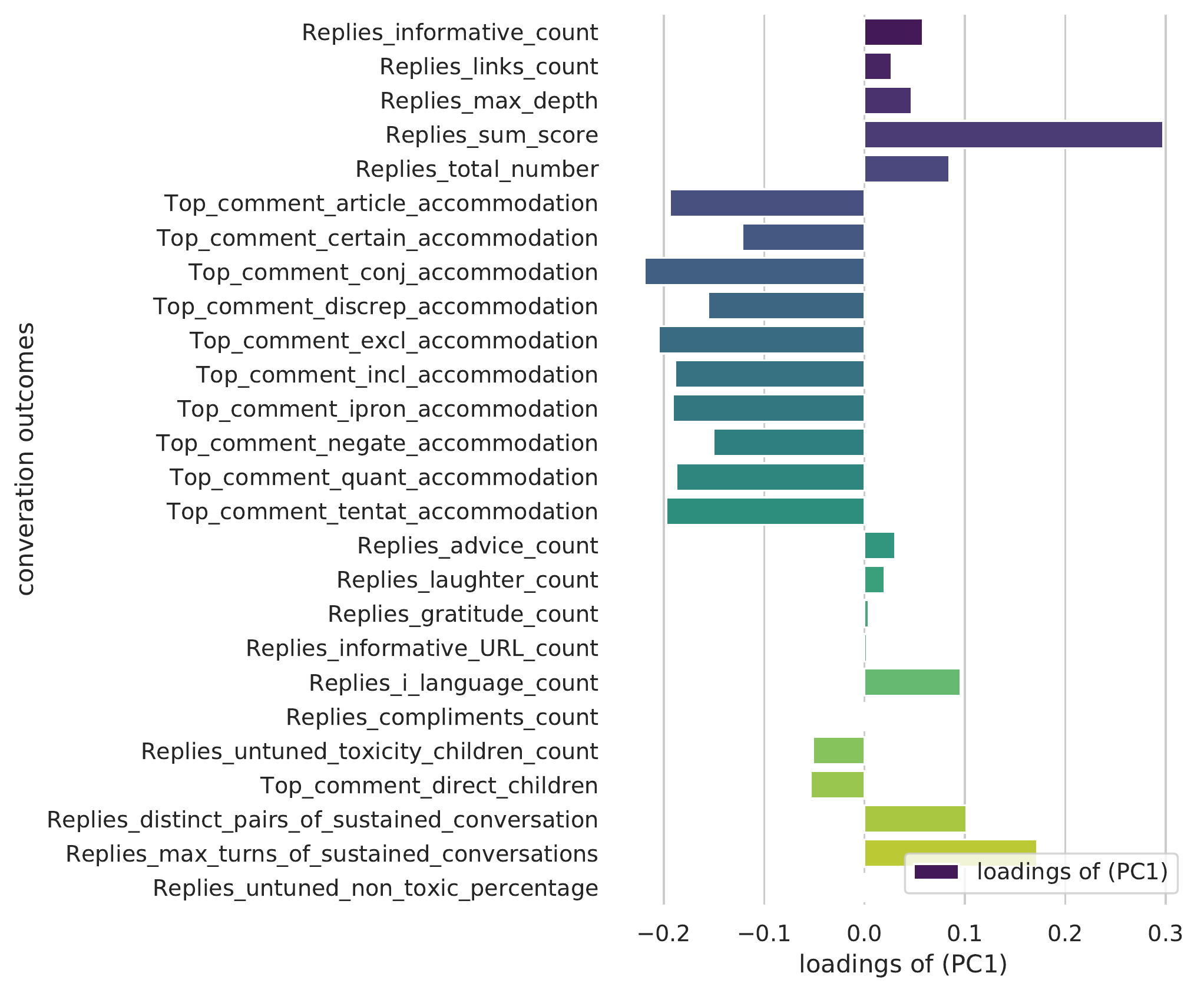}
    \caption{PCA Component 1 Loadings Across Metrics.}
    \label{fig:loadingforpc1}
\end{figure}

\subsection{Politeness}
\label{appendix-politeness}
Two prior datasets exist with politeness ratings. The data of \citet{danescu2013computational} contains z-scored ratings of politeness for questions, whereas the data for \citet{wang2018its} contains ratings for statements of a variety of lengths rated on a scale in [1,5] where 3 indicates neither polite nor impolite. To build a robust classifier for Reddit, we combine both datasets, and rescale both datasets to be within [-1,1]. 
To obtain the politeness regressor, we first pre-train a BERT-based model \cite{devlin2018bert} on Reddit data using masked language modeling. Then, we fine-tune those parameters using the Adam optimizer with a learning rate of 0.00002. The max sequence length is set to 128. While training, we adopt $MSE$ as the loss function and a five-fold cross validation strategy is utilized when evaluating model's performance. Each model was run at most 5 epochs and we took the one whose average Pearson $r$  across five folds was the highest for further usage. The final model obtained a $r$=0.66 with human judgments from both datasets.

\subsection{Supportiveness}
\label{app:support}
The supportiveness regressor was trained in a similar manner as the politeness regressor, but used the only available dataset of  \citet{wang2018its} for estimating support. Support is scored within [-1,1] with a rating of 0 indicating neither supportive nor unsupportive. A BERT model is first pre-trained on  Reddit data using masked language modeling and then five-fold cross-validation is done where each fold is fine-tuned on these support ratings. We select the model with the highest rating across folds. The final model reached $r$=0.58 with human judgments in their data, which surpasses the state-of-the-art model results reported in  \citet{wang2018its} for their best model.

\section{Additional PCA Analysis}

Multiple prosocial behaviors may occur in the same conversation and to capture their regular co-occurrence, we use Principal Component Analysis (PCA) to identify the main forms of variation. PCA is computed on a matrix where each conversation is a row and the columns contain the value of each prosocial metric Shown in Figure \ref{fig:explained-variance-from-pca} (main paper), the first principle component explains 57.4\% of the variance in the data, with all other components explaining far less. The loadings of this first principle component (Figure~\ref{fig:loadingforpc0})  shows that this component is loading on all of the prosocial behaviors (and negatively on the antisocial behaviors) indicating that it effectively summarizes our studied prosocial behaviors within a single metric. As a comparison, we show loadings for the second largest component in Figure \ref{fig:loadingforpc1}, which explains $\sim$10\% of the variance; this component does not have any clear association with prosocial behavior and seems to match conversations with high scores but little conversation.
Similar trends were observed for all other components, which lacked a clear association with prosocial behavior, suggesting that a single metric can be a reasonable proxy for summarizing the prosocial behaviors.

\begin{table}[tb]
  \centering
  \resizebox{0.49\textwidth}{!}{
    \begin{tabular}{r c c | r c c }
                   & Our Model & XGBoost  & & Our Model & XGBoost \\
\hline                   
        Art        & 1.88      & 1.75    &
        Pictures   & 2.05      & 1.93    \\
        Culture    & 2.14      & 2.01    &
        Music      & 2.00      & 2.05    \\
        TV         & 2.16      & 2.09    &
        Lifestyle  & 2.16      & 2.19    \\
        Sports     & 2.27      & 2.19    &
        Movies     & 2.50      & 2.29    \\
        Gaming     & 2.32      & 2.30    &
        Humor      & 2.54      & 2.39    \\
        Technology & 2.55      & 2.50    &
        Meta       & 2.77      & 2.59    \\
        Discussion & 2.70      & 2.62    &
        Location   & 2.70      & 2.63    \\
        Info       & 2.85      & 2.79    &
        Science    & 3.07      & 2.97    \\
    \end{tabular}
  }
  \caption{The MSE of prosocial forecasts within different subreddit categories shows  that our two top models both attain higher performance in communities whose discussion relates to pop-culture such as Movies, Art, and Culture.}
  \label{tab:category-mse-loss}%
\end{table}%

\begin{table}[tb]
  \centering
    \resizebox{0.43\textwidth}{!}{
    \begin{tabular}{r c}
        Hyperparameter & Value \\
        \hline
        booster type & gbtree \\
        learning rate $\eta$                     & 0.05 \\
        minimum loss reduction $\gamma$                   & 1.0  \\
        maximum depth of a tree              & 4    \\
        minimum child weight        & 1.0  \\
        subsample ratio of training instances               & 0.8  \\
        the subsample ratio of columns (colsample\_bytree)       & 0.8  \\
        L2 regularization term on weights $\lambda$                  & 3.0  \\
        L1 regularization term on weights $\alpha$                   & 1.0  \\
        number of parallel trees     & 1    \\
        number of boost rounds       & 5000 \\
        number of early stopping rounds & 50  
        \end{tabular}
    }
  \caption{Hyperparameters of the XGBoost Model}
  \label{tab:xgboost:hyperparameters}%
\end{table}%

\begin{table}[tb]
  \centering
    \resizebox{0.45\textwidth}{!}{
    \begin{tabular}{r c c }
        Hyperparameter & Our Model & Frozen Albert \\
        \hline
        learning rate                      & 1e-5 & 1e-4\\
        number of epochs                  & 2  & 5 \\
        L2 penalty (c)              & 1e-6   &  1e-6 \\
        pretrained albert type (for \tlc texts)      & albert-base-v2 & albert-base-v2  \\
        pretrained albert type (for post texts)      & albert-base-v2 & albert-base-v2 \\
        dropping probability of the dropout layer           & 0.5 & 0.5 \\
        subreddit embedding dimension                  & 16 & 16\\
        learning rate scheduling & linear & linear\\
        optimizer & AdamW & AdamW \\
        random seed & 42 & 42 \\
        \end{tabular}
    }
  \caption{Hyperparameters of our model (left) and the model using frozen weights for the base Albert Model (right).}
  \label{tab:ourmodel:hyperparameters}%
\end{table}%

\section{Model Hyperparameters}
\label{app:hyperparameters}

\paragraph{XGBoost}
Hyperparameters for the XGBoost model are shown in Table \ref{tab:xgboost:hyperparameters}. We tuned the learning rate $\eta$ through grid search (log-linearly) in the ranges between 0.1 and 0.001. We regard models that have the lowest mean square loss as the best model. The model was trained on cpu-s for 27 hours, 21 minutes and 3 seconds, validated on the validation set every 10 iterations. The mean square error of the above model on the validation set is 1.49, and its $R^2$ is 0.27.

\paragraph{Our Albert-based Model}
Hyperparameters for our Albert-based model are shown in Table \ref{tab:ourmodel:hyperparameters}. We tuned the learning rate and weight decay using grid search (log-linearly) in the ranges from 0.1 to 0.001, and from 1e-4 to 1e-7 respectively. We regarded models that have the lowest mean square loss as the best model. The model was trained on a single GeForce GTX 2080 Ti device for 45 hours, 53 minutes and 14 seconds, validated on a randomly-sampled validation set every 3351 iterations (40 times per epoch). The mean square error of the above model on the validation set is 2.24, and its $R^2$ is 0.27.
For the model using frozen Albert-parameters from the Hugging Face transformer library \cite{wolf-etal-2020-transformers}, parameters are also shown in Table~\ref{tab:ourmodel:hyperparameters}. This model has the same architecture, but differs only in fine-tuning and which weights are frozen.
The model was trained on a single GeForce GTX 2080 Ti device for 55 hours, 34 minutes and 9 seconds, validated on a randomly-sampled validation set every 482 iterations (40 times per epoch). The mean square error of the above model on the validation set is 2.50, and its $R^2$ is 0.16.

\end{document}